\documentclass[aps,prb,10pt,twocolumn,showpacs,superscriptaddress]{revtex4-1}
\usepackage{amssymb,graphicx,color,amsmath,xfrac,bm,stmaryrd,trimclip, array}

\usepackage[mathletters]{ucs}
\usepackage[utf8x]{inputenc}
\definecolor{lightblue}{rgb}{0.17,0.39,1}
\usepackage[bookmarks, colorlinks=true, breaklinks]{hyperref}
\hypersetup{linkcolor=lightblue, citecolor=lightblue, filecolor=black, urlcolor=lightblue} 

\newcommand{\fra}[2]{ {#1}/{#2}}

\newcommand{\para}[1]{\ensuremath{\left(#1\right)}}
\newcommand{\bigpara}[1]{\ensuremath{\Big(#1\Big)}}
\newcommand{\brac}[1]{\ensuremath{\left[#1\right]}}
\newcommand{\curly}[1]{\ensuremath{\left\{#1\right\}}}
\newcommand{\av}[1]{\ensuremath{\left\langle{\!{#1}\!}\right\rangle}}
\newcommand{\avav}[1]{\av{\av{#1}}}
\newcommand{\bra}[1]{\ensuremath{\left\langle{{#1}}\right|}}
\newcommand{\ket}[1]{\ensuremath{\left|{{#1}}\right\rangle}}

\newcommand{\der}[2]{\para{\frac{∂{#1}}{∂{#2}}}}

\newcommand{\lap}{∇^2}
\newcommand{\grad}{∇}
\renewcommand{\Im}{{\mathrm{Im}}}
\renewcommand{\Re}{{\mathrm{Re}}}

\newcommand{\matr}[1]{\para{    \begin{matrix}#1    \end{matrix}}}
\newcommand{\qmatr}[1]{\brac{     \begin{matrix}#1   \end{matrix}}}

\makeatletter
\renewcommand*\env@matrix[1][c]{\hskip -\arraycolsep
  \let\@ifnextchar\new@ifnextchar
  \array{*\c@MaxMatrixCols #1}}
\makeatother

\newcommand{\bmatr}[1]{{  \begin{pmatrix}[r]#1  \end{pmatrix}}}

\newcolumntype{C}[1]{>{\centering\arraybackslash$}m{#1}<{$}}
\newlength{\micolwd}           
\newcommand{\smatr}[1]{\para{\begin{smallmatrix} #1  \end{smallmatrix}} }

\newcommand{\flux}{ {\ensuremath{\text{flux}}}}
\newcommand{\ext}{ {\ensuremath{\text{ext}}}}
\newcommand{\Tr}{ {\ensuremath{\text{Tr}\,}}}
\newcommand{\n}{ {\ensuremath{\bm{n}}}}

\newcommand{\B}{ {\ensuremath{\bm{B}}}}
\newcommand{\M}{ {\ensuremath{\bm{M}}}}
\newcommand{\g}{ {\ensuremath{\hat{g}}}}
\renewcommand{\S}{ {\ensuremath{\hat{S}}}}
\newcommand{\N}{ {\ensuremath{\hat{N}}}}
\newcommand{\mi}{ {\ensuremath{ \av{x^2} }}}
\newcommand{\St}{ {\ensuremath{S}}}
\newcommand{\Lever}{ {\ensuremath{\rm{L}}}}
\newcommand{\Sample}{ {\ensuremath{\rm{S}}}}
\newcommand{\Bohr}{ {\ensuremath{\rm{B}}}}
\newcommand{\Boltz}{ {\ensuremath{\rm{B}}}}
\newcommand{\Boltzmann}{ {\ensuremath{\rm{B}}}}
\newcommand{\iso}{ {\ensuremath{\rm{iso}}}}

\newcommand{\To}{{\ensuremath{T}}}
\newcommand{\Top}{{\ensuremath{\mathcal{T}}}}
\newcommand{\Mop}{{\ensuremath{\mathcal{M}}}}

\begin{document}
\title{Magnetotropic susceptibility.}

\author{A.~Shekhter}
\email[Email: ]{arkadyshekhter@gmail.com}
\affiliation{Los Alamos National Laboratory, Los Alamos NM 87545 USA}
\author{R.~D.~McDonald}
\affiliation{Los Alamos National Laboratory, Los Alamos NM 87545 USA}
\author{B.~J.~Ramshaw}
\affiliation{Laboratory of Atomic and Solid State Physics, Cornell University, Ithaca NY 14853 USA}
\author{K.~A.~Modic}
\affiliation{Institute of Science and Technology Austria, Am Campus 1, 3400 Klosterneuburg Austria} 
					
	\begin{abstract}  
The magnetotropic susceptibility is the thermodynamic coefficient associated with the rotational anisotropy of the free energy in an external magnetic field, and is closely related to the magnetic susceptibility. It emerges naturally in frequency-shift measurements of oscillating mechanical cantilevers, which are becoming an increasingly important tool in the quantitative study of the thermodynamics of modern condensed matter systems. Here we discuss the basic properties of the magnetotropic susceptibility as they relate to the experimental aspects of frequency-shift measurements, as well as to the interpretation of those experiments in terms of the intrinsic properties of the system under study. 
	\end{abstract}

\date{\today}\maketitle 

\setlength{\parindent}{0.5cm}
\setlength{\parskip}{0.2cm}


\section{Introduction.}
\label{sec:intro}

High-$Q$ mechanical oscillators have long been used in the study of thermodynamic properties of condensed matter systems. In Refs.~\onlinecite{Bishop1978, Kleiman1984, Kleiman1985, Kleiman1987, Worthington1987, Bishop1992, Bolle1999, Chiaverini2001}, for example, oscillations of a sample in a magnetic field are used to study the thermodynamic behavior of a vortex lattice. Recent advances in lithographic techniques have expanded the use of high-$Q$ mechanical oscillators in studying condensed matter systems other than superconductors, such as spin-liquids, \cite{Modic2018a, Modic2021} correlated metals, \cite{Pocs2021, Mumford2021} and unconventional superconductors.\cite{Maeno2011} In these experiments, the physical properties of the sample are inferred from the shift of the resonance frequency of the cantilever-sample assembly. Qualitative insight into the behavior of the physical systems in this broader scientific context requires a quantitative interpretation of frequency shifts. \cite{Maeno2011,Mumford2021, Modic2021} 

A frequency shift in a mechanical oscillator can be induced by the oscillating linear motion of a sample in a non-uniform magnetic field or by the oscillating rotational motion in a uniform applied magnetic field. Small frequency shifts are also accompanied by resonance width broadening associated with relaxation phenomena in the sample coupled to either the  rotational- or linear-oscillating motion of the sample. 

In this paper, we focus on the {\it magnetotropic susceptibility} that captures the changes in the free energy of a magnetically anisotropic sample associated with its rotation in a uniform magnetic field. The frequency shift of the resonance of the cantilever-sample assembly in a uniform external magnetic field is captured entirely by the magnetotropic susceptibility of the sample. Dynamic magnetotropic susceptibility captures the relaxation phenomena coupled to the rotation of the sample in an applied magnetic field. 

In Section~\ref{sec:magsusc}, we introduce the magnetotropic susceptibility in a rigorous manner, independent of the measurement technique. We define the magnetotropic susceptibility as a thermodynamic correlation function. After that, we discuss the dynamic, or frequency-dependent, magnetotropic susceptibility which encapsulates relaxation phenomena in the sample induced by its rotation in an applied magnetic field. Dynamic magnetotropic susceptibility is accessed directly in the measurements of the width of the resonance of the cantilever induced by the relaxation in the sample. 

In Section~\ref{sec:frequencyshiftmeas}, we discuss the emergence of the magnetotropic susceptibility in frequency-shift measurements and how certain experimental aspects relate to the intrinsic properties of the sample under study. 
Finally, in Section~\ref{sec:cantilever}, we consider the mechanics of the cantilever in the thin-plate approximation. This section presents a self-contained discussion of the bending stiffness of the cantilever, which determines the sensitivity of frequency shift measurements of the magnetotropic susceptibility. 


	\begin{figure}[t!!]
				\centerline{  
				\includegraphics[width=0.7\columnwidth]{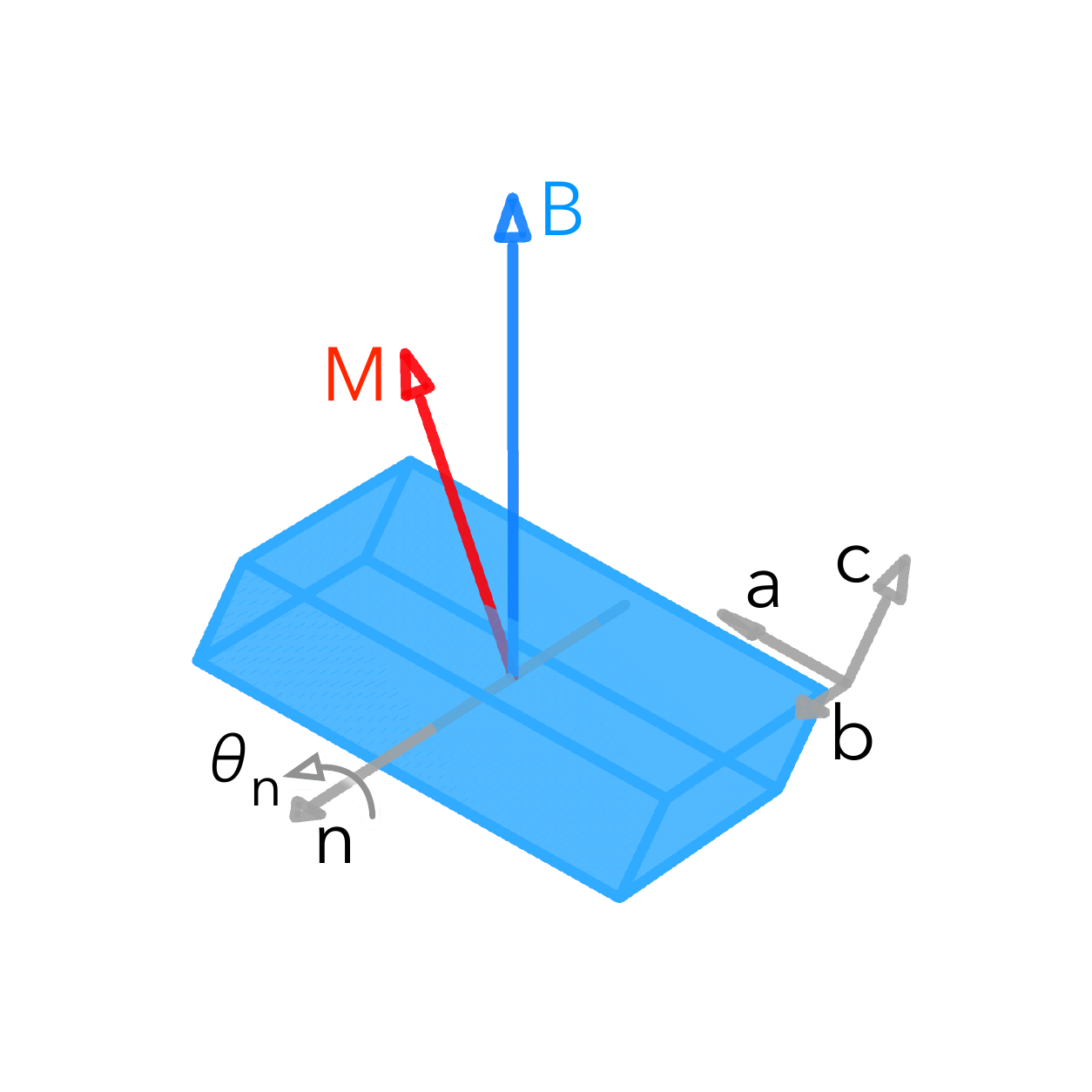}  }
				\caption{ The sample is rotated in an external magnetic field $\B$ around axis $\n$. The magnetotropic susceptibility $k_{\n}$ is the curvature---the second derivative---of the angular dependence of free energy in the applied magnetic field. $\n$ indicates a slice of rotations by angle $θ_{\n}$ around axis $\n$. 
				}
				\label{fig:fig1} 
				\end{figure}

\section{The magnetotropic susceptibility. }
\label{sec:magsusc}

\subsection{Definition and basic properties.}
\label{sec:definition}

The free energy of a magnetically anisotropic sample depends on the direction of the magnetic field with respect to the crystallographic directions of the sample. One can study a ``slice'' of the overall angular dependence of the free energy, $F(θ_{\n}, \B)$, where the magnetic field  $\B$ rotates around axis $\n$ without change in its magnitude. The angular variable $θ_{\n}$ considered as a thermodynamic parameter defines the magnetic torque, $\To_{\n}(\B)$~:
		\begin{align}\label{eq:torque-def}
dF(\B) = \To_{\n}(\B) \; dθ_{\n} \,, \quad   \To_{\n}(\B) =  - \M ⋅ (\n×\B) \,.  
		\end{align} 
Magnetic torque is determined by the magnetization $\M  =  - dF/d\B$ induced by a small change in magnetic field $δ_{dθ\n} \B =  ( \n × \B ) \, dθ_{\n} $ that is incurred by a small rotation $dθ_{\n}$ around axis $\n$  (Figure~\ref{fig:fig1}). 
 
The diagonal thermodynamic coefficient \cite{Callen1985}  associated with the thermodynamic parameter $θ_{\n}$ is the magnetotropic susceptibility, 
		\begin{align}\label{eq:fromfree1}
& d\To_{\n}(\B)  =  k_{\n}(\B) \; dθ_{\n} \,, \notag\\ 
&	 k_{\n}(\B)  \!=\!  (\n\!×\!\B) \!⋅\! (\n\!×\!\M) \!-\! (\n\!×\!\B) \!⋅\!\hat{χ}(\B) \!⋅\! (\n\!×\!\B)	\,,
		\end{align}
where $\hat{χ}_{μν}(\B) = - \fra{dM_{μ}}{dB_{ν}}$ is the magnetic susceptibility tensor. 
The two terms in the second line of Eq.~(\ref{eq:fromfree1}) have a different physical character, which will become apparent in the discussion of correlation functions in Sections \ref{sec:corrfunc-thermo} and \ref{sec:corr-func-dynamic}. The explicit dependence of the magnetic torque $\To_{\n}(\B) =  -\M ⋅  (\n ×\B )$  on the rotation of magnetic field produces the first term in Eq.~(\ref{eq:fromfree1}), $ -\M ⋅  (\n × (\n ×\B ) )$. The implicit dependence of torque on magnetic field through changes in the magnetization produces the second term. 


It will be convenient to rewrite Eq.~(\ref{eq:fromfree1}) in a matrix form using generators of rotation in vector representation, $i(\S_{μ})_{ij}= ϵ_{μ ij}$, satisfying the commutation relations $ [ \S_{i} , \S_{j} ] = i ϵ_{ijk} \S_{k} $ where  $ [ \S_{i} , \S_{j} ] ≡  \S_{i} ⋅ \S_{j} - \S_{j} ⋅ \S_{i}$, 
\begin{align} \label{eq:O3-generators}
&		\S_1 \!=\! i\smatr{ 
0&   0&  0\\ 
0&    0& -1\\ 
0&    1&  0}, \, 
\S_2 \!=\! i\smatr{ 
0&  0&  1\\ 
0&  0&  0\\ 
-1&  0&  0}, \,
\S_3 \!=\!  i\smatr{ 
0&  -1& 0\\ 
1&  0& 0\\ 
0&  0&  0} \,.
\end{align}  
Rotation of a vector represented by the cross product $ \n × \B$ in Eq.~(\ref{eq:fromfree1}) can be represented equivalently by a matrix multiplication 
\begin{align} \label{eq:cross-matrix}
& δ_{θ\n} \bm{B} = (\n × \B) dθ  =  (i\N\!⋅\!\B ) dθ \,, \notag\\
	&\quad  i\N = -i\S_{μ}n_{μ} =  \qmatr{ 
0    &  -n_z  & n_y\\ 
n_z &  0    & -n_x\\ 
-n_y  & n_x  &  0} \,.
\end{align}  
 In this representation, the magnetotropic susceptibility takes the form 
		\begin{align}\label{eq:fromfree3}
	k_{\n}(\B)  \!=\!  \B\!⋅\! \N \!⋅\! 
	 \N\!⋅\! \M - \B\!⋅\! \N \!⋅\!\hat{χ} \!⋅\! \N  \!⋅\! \B 	\,.
		\end{align}

The magnetotropic susceptibility $k_{\n}(\B)$, as well as the magnetic torque $\To_{\n}(\B)$, describes only a slice of the complete angular dependence of the free energy. 
For arbitrary rotation slice $\n$, magnetic torque is specified by a scalar product of an axial vector $\bm{\To}= \M×\B$ and vector $\n$,  via $\To_{\n}(\B) =  \bm{\To}_μ\n_μ  $. 
Equations~(\ref{eq:fromfree1}) and (\ref{eq:fromfree3}) define the magnetotropic susceptibility $k_{\n}(\B) $ as a bilinear function of the components of the vector $\n$.  The generalization of the torque vector $\bm{\To}$ to the magnetotropic susceptibility is a symmetric second-rank tensor $k_{μν}(\B)$, 
\begin{align}\label{eq:definition}
		k_{\n}(\B) = &  k_{μν}(\B) n_{ν} n_{μ} \,. 
	\end{align} 
The magnetotropic susceptibility tensor $k_{μν}(\B)$ encodes the second angular derivative of the free energy for all slice directions $\n$, just like the torque vector $\bm{\To}$ encodes its first derivative. Tensor components of $k_{μν}(\B)$ follow from Eqs. (\ref{eq:cross-matrix}) and (\ref{eq:fromfree3}), with subsequent symmetrization~:
		\begin{align}\label{eq:definition1}
	k_{μν}(\B)  = & \, \frac12 \B\!⋅\! ( \S_μ \!⋅\! 
	 \S_ν + \S_ν\!⋅\! \S_μ ) \!⋅\! \M  \notag\\
	 & -  \, \frac12 \B\!⋅\! ( \S_μ \!⋅\!\hat{χ} \!⋅\! \S_ν + \S_ν \!⋅\!\hat{χ} \!⋅\! \S_μ )  \!⋅\! \B 	\,.
		\end{align}
The first line can also be written as $ δ_{μν} (\B \!⋅\! \M) 	- \sfrac12(B_{μ} M_{ν} + B_{ν} M_{μ})$. 

The product $\N \!⋅\! \B$ in Eq.~(\ref{eq:fromfree3}) as well as the cross-product $(\n\!×\!\B)$ in Eq.~(\ref{eq:fromfree1}) vanish for $\n$ along the magnetic field $\B$. Therefore, both magnetic torque and magnetotropic susceptibility vanish. This reflects the physical fact that rotations of a sample around axis $\n$ along magnetic field $\B$ have no effect on the magnetic part of the free energy. Therefore, the product $k_{μν}(\B)n_μ n_ν$  vanishes for $\n$ along $\B$, 
	\begin{align}\label{eq:transverse}
k_{μν}(\B) \B_{μ}\B_{ν} ≡ 0\,,
	\end{align}
because $(\S_{μ}\B_{μ})\!⋅\! \B ≡ 0$. This is analogous to the vanishing of the scalar product of the torque vector and the magnetic field, $\bm{\To}_{μ}\B_{μ} ≡ 0$. For torque, this identity means that the magnetic torque vector lives in a plane perpendicular to vector $\B$. Similarly, the six components of the symmetric magnetotropic susceptibility tensor $k_{μν}(\B)$ are constrained by one condition, Eq.~(\ref{eq:transverse}).


In the linear regime, the magnetotropic susceptibility is bilinear in the components of magnetic field $\B$. The magnetization $\M = \hat{χ}^0\!⋅\! \B$ is linear in magnetic field and linear magnetic susceptibility $\hat{χ}^0$ is independent of magnetic field. Equation~(\ref{eq:fromfree3}) reduces to 
		\begin{align}\label{eq:fromfree2} 
&	k_{\n}^0(\B)  \!=\!  \B\!⋅\! \N \!⋅\! 
	[ \N ,  \hat{χ}^0 ]  \!⋅\! \B 	\,,
		\end{align}
	and Eq.~(\ref{eq:definition1}) to 
		\begin{align}\label{eq:definition1-linear}
	k_{μν}^0(\B)  = & \, \frac12 \B\!⋅\! ( 
	\S_μ \!⋅\! [\S_ν , \hat{χ}^0 ] 
+	\S_ν \!⋅\! [\S_μ , \hat{χ}^0 ]  	)  \!⋅\! \B 	\,.
		\end{align}
If we choose the axes $x, y, z$ along the crystallographic directions $a, b, c$ (Figure \ref{fig:fig1})---or the principal directions of magnetic susceptibility ${χ}$ when the crystal symmetry is lower than orthorhombic---then the linear magnetic susceptibility tensor $\hat{χ}^0$ is diagonal  $\hat{χ}^0=\text{diag}\{χ^0_{xx}, χ^0_{yy}, χ^0_{zz} \}$. The full magnetotropic tensor,  Eq.~(\ref{eq:definition1-linear}), in this basis has the form, 
\begin{widetext}
\begin{align}
	k_{μν}^0(\B)  = \bmatr{
	(B_y^2-B_z^2)(  χ^0_{yy} - χ^0_{zz} ) 
	& - B_xB_y (\sfrac{χ^0_{xx}}{2} + \sfrac{χ^0_{yy}}2 - χ^0_{zz}) 
	& - B_xB_z (\sfrac{χ^0_{xx}}2 - χ^0_{yy} + \sfrac{χ^0_{zz}}2 ) 
	\\
	 - B_xB_y (\sfrac{χ^0_{xx}}2 + \sfrac{χ^0_{yy}}2 - χ^0_{zz})
	& (B_x^2-B_z^2)(  χ^0_{xx} - χ^0_{zz} ) 
	& B_yB_z ( χ^0_{xx} - \sfrac{χ^0_{yy}}2-\sfrac{χ^0_{zz}}2) 
	\\
	 - B_xB_z  (\sfrac{χ^0_{xx}}2 - χ^0_{yy} + \sfrac{χ^0_{zz}}2 )
	& B_yB_z ( χ^0_{xx} - \sfrac{χ^0_{yy}}2-\sfrac{χ^0_{zz}}2) 
	& (B_x^2-B_y^2)(  χ^0_{xx} - χ^0_{yy} ) 
	} \,.
\end{align}
\end{widetext}
Note that only two combinations of the components of the linear magnetic susceptibility in this expression are linearly independent. Therefore,  the linear magnetotropic tensor $k_{μν}^0(\B)$ determines two (out of three) independent components of linear magnetic susceptibility, consistent with the more general discussion around  Eq.~(\ref{eq:transverse}).

For example, a slice of rotations around the $\n=\hat{y}$ axis (parallel to $b$) is described with $\N = -\S_{2}$ in Eq.~(\ref{eq:fromfree2})~: 
\begin{align} 
 k_{\hat{y}}^0(\B)  = & \B	\!⋅\!  \S_{2} \!⋅\! [ \S_{2} ,  \hat{χ}^0 ] \!⋅\! \B \notag\\
	=&   \B \!⋅\! \qmatr{ 
 	χ_{xx}^0-χ_{zz}^0 	&	0	&	0\\ 
	0 	&	0	&	0	\\ 
	0  	&	0	&	-(χ_{xx}^0-χ_{zz}^0)} \!⋅\!  \B \,. 
\end{align}
Note the vanishing eigenvalue corresponding to direction $\hat{y}$. 
In components, 
	\begin{align} \label{eq:shminear}
 k_{\hat{y}}^0(\B) 
 \!=\!  (B_c^2\!-\!B_a^2)\,(χ _c^0\!-\!χ_a^0)  =  B_{ac}^2\, (χ_c^0\!-\!χ_a^0) \cos\!2θ \,, 
	\end{align}
where $B_{ac}=(B_a^2+B_c^2)^{1/2}$ is the component of magnetic field $\B$ in the $ac$-plane and $θ$ is the angle between $\B_{ac}$ and the $c$-axis, $(B_a,B_c)= B_{ac}(\sinθ, \cosθ)$. 
The linear magnetotropic coefficient $k_{\hat{y}}(\B)$ has the same angular dependence in the $ac$-plane as the free energy in the linear regime, $F^0(\B) = (1/2) {χ}^0_{ij}B_iB_j = - (1/4)( {χ}^0_{c}-{χ}^0_{a}) B_{ac}^2 \cos2θ$. The magnetotropic susceptibility does not vanish for applied magnetic fields along crystallographic directions. 
Equation~(\ref{eq:shminear}) also shows that the magnetotropic susceptibility is positive for magnetic field along the easy axis, where the free energy is near its local minimum. 


For completeness, we can also consider the magnetotropic susceptibility of a polycrystalline sample. The free energy of a polycrystalline sample is isotropic,  $F(\B) = F_{\iso}(|\B|)$, where $F_{\iso}(|\B|)$ is the average of the anisotropic free energy over all crystal lattice orientations. 
The magnetotropic susceptibility of a polycrystal is determined by the angular derivatives of $F_{\iso}(|\B|)$ and therefore is zero. 
When considering Eq.~(\ref{eq:fromfree1}) or Eq.~(\ref{eq:fromfree3}) in a polycrystalline sample, the vanishing of the magnetotropic susceptibility is not immediately clear because it arises as a cancellation between the first and second terms. Such cancellation is obvious in the linear regime, Eq.~(\ref{eq:fromfree2}), where one has to replace the linear magnetic susceptibility  $\hat{χ}^0$ with its polycrystalline average, proportional to the unit matrix. 

In the non-linear regime, the character of the first and second terms in Eq.~(\ref{eq:fromfree1}) and  Eq.~(\ref{eq:fromfree3}) is quite different,  and the cancellation is not evident. It is instructive, though, to track the cancellation starting from a non-linear magnetization and a non-linear magnetic susceptibility, 
	\begin{align}
& \M=  -\frac{dF}{d\B}   =  -\frac1{B} \frac{dF}{dB}  \B \,,	\\ 
& 		{χ}_{μν} =   \frac{d\M_μ}{d\B_ν} 
			 = \para{ \frac1{B} \frac{d(M/B)}{dB} } \B_μ\B _ν 
				+\frac{M}{B} δ_{μν} \,,
\label{eq:nonlinear-chi}
	\end{align}
where $B=|\B|$ and $M=|\M|$. 
The first term in Eq.~(\ref{eq:nonlinear-chi}) is proportional to the projector matrix in the direction of the magnetic field. Therefore, the non-linear magnetic susceptibility ${χ}_{μν}$ of a polycrystalline sample is a uniaxial tensor with magnetic field as a symmetry axis. This projector part of the magnetic susceptibility is projected out when substituted into  Eq.~(\ref{eq:fromfree1}), and, therefore, does not affect the magnetotropic susceptibility. 
The second term in Eq.~(\ref{eq:nonlinear-chi}) is isotropic.  When substituted into Eq.~(\ref{eq:fromfree1}), it cancels with the first term in Eq.~(\ref{eq:fromfree1}). 
    
It will be shown in Section \ref{sec:corr-func-dynamic} that the relaxation phenomena induced by the rotation in an applied magnetic field are captured by the imaginary part of the dynamic (frequency dependent) magnetotropic susceptibility. In the polycrystalline sample, these dissipative phenomena are induced independently in each single-crystal grain and their total effect on the dynamic magnetotropic susceptibility does not add up to zero. Therefore, the polycrystalline sample has a non-zero imaginary part of the dynamic magnetotropic susceptibility (Section \ref{sec:corr-func-dynamic}) and, by analyticity, a non-zero real part at finite frequency.    
   
\subsection{Example: An isolated spin-$\sfrac12$ }
\label{sec:spin-one-half}

\begin{figure}[ht!!]
				\centerline{  
				\includegraphics[width=0.85\columnwidth]{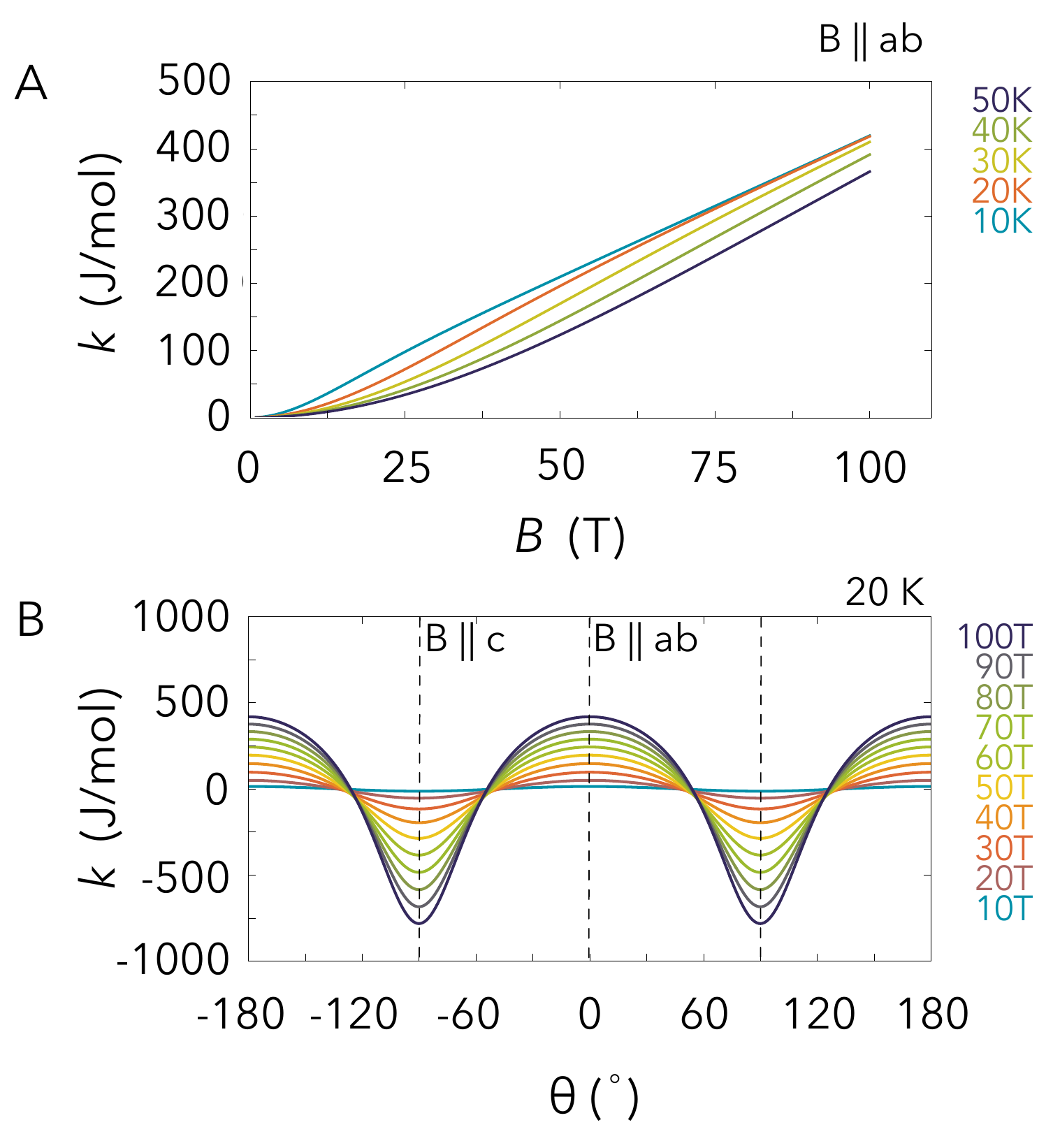}  }
				\caption{\scriptsize The magnetotropic susceptibility calculated for an isolated spin-$\sfrac12$ with an anisotropic $g$-factor, $g_{a,b}=2, g_c=1$. The rotation axis is along the $b$-axis of the crystal lattice. 
				A. Field scans up to 100 T. Magnetic field is along the $a$-axis. 
				B. Angular scans at 20 K. Easy axis is along the $ab$-plane. The hard axis is along the $c$-axis.  
				}
				\label{fig:fig2} 
				\end{figure}

As a simple example, here we briefly discuss the magnetotropic susceptibility of an isolated spin-$\sfrac12$ with an anisotropic $g$-factor, described with Hamiltonian
							\begin{align}\label{eq:torque-def-spin}
H_0 = μ_{\Bohr} \B \!⋅\!  \g \!⋅\! (\bm{σ}/2)\,, 
							\end{align}
where $\bm{σ} = \curly{σ_j}$ are three Pauli matrices and $\g=g_{ij}$ is a symmetric $g$-factor tensor. The free energy can be calculated in a closed form,
		\begin{align}\label{eq:torque-defenergy-spin12}
		&e^{-\frac{F(\B)}{T}} =  \Tr e^{-\frac{H_0}{T}}    \,, \notag\\
		&\quad  F(\B) =  f(a)	=  -  T\log\para{  2\!\cosh\! \frac{ \sqrt{ a } }{2T} } \,, \notag\\
		& \quad\quad a = μ_{\Bohr}^2 \, \B ⋅   \g  ⋅ \g^{⊤}  ⋅ \B  \,,
		\end{align}
where $\Tr ⋯ $ stands for a trace of 2x2 matrices and $T$ is temperature measured in energy units. In the basis $a,b,c$ where the $g$-factor tensor $\g$ is diagonal, $a =  μ_{\Bohr}^2 \, \para{ g_{aa}^2B_a^2 + g_{bb}^2 B_b^2 + g_{cc}^2 B_c^2 }$,   
the torque $\To_{\n}$ and the magnetotropic susceptibility $k_{\n}$ are 
		\begin{align}
& 		\To_{\n} =    \frac{df(a)}{dθ} 
		=  \frac{df(a)}{da} × {\frac{da}{dθ}}\,, \notag\\ 
& 	 	k_{\n} =\frac{d^2f(a)}{dθ^2} 
		=  \frac{d^2f(a)}{da^2}  ×  \para{\frac{da}{dθ}}^2 
			+  \frac{df(a)}{da} × \frac{d^2a}{dθ^2}   \,,
		\end{align}
where the angular derivatives of magnetic field vector are evaluated using Eq.~(\ref{eq:cross-matrix}). We have
							\begin{align}
&\frac{da}{dθ}           =   μ_{\Bohr}^2  \B ⋅  [ \g ⋅ \g^{⊤} ,  i\N ] ⋅ \B  \,, \notag\\
&\frac{d^2a}{dθ^2} = μ_{\Bohr}^2   \B⋅ [ [ \g ⋅ \g^{⊤}, i\N  ], i\N  ] ⋅ \B   \,. 
							\end{align}
Alternatively, one can use Eqs.~\ref{eq:torque-def} and \ref{eq:fromfree1} with the magnetization and magnetic susceptibility of an isolated spin $\sfrac12$ given by   
							\begin{align}
	\M = &  - \frac{df(a)}{da}  ×  \frac{da}{d\B} 
		= 	μ_{\Bohr}^2  \frac{df(a)}{da}\,  ×  (\!-2\,\g\!⋅\!\g^{⊤}\!\!⋅\!\B )		\,,					\notag\\
  	\hat{χ}_{μν} =& \frac{d\M_μ}{d\B_ν}  
  		=  	μ_{\Bohr}^2 \frac{df(a)}{da}  \, (\!-2\, \g \!⋅\! \g^{⊤} )_{μν}   
			- \frac{ \fra{d^2f(a)}{da^2} }{ \para{\fra{df(a)}{da}}^{\!2}} \, \M_{μ}\M_{ν} \,.
							\end{align}
Figure~\ref{fig:fig2} shows the angular- and magnetic field dependence of the magnetotropic coefficient of an isolated spin-$\sfrac12$ with an anisotropic $g$-factor. 
As discussed in Section \ref{sec:examples}, for a 0.1~nanomol size sample of spin-1/2 system, the shift of the frequency of the fundamental resonance mode at 50~kHz of a 180~nJ bending stiffness cantilever is a few kilohertz at 50~K and 50~T (Figure~\ref{fig:fig2}A), much larger than the resonance width of a fraction of a hertz. The frequency shifts of a millihertz and smaller are readily measured by standard locking techniques. 

A non-interacting spin-$\sfrac12$ system provides the simplest example of a scale-invariant behavior---the scale and the character of the magnetic-field dependence are determined by the temperature alone. In particular, at high magnetic fields, $\mu_{\Bohr} B \gg k_{\Boltz} T$, the free energy is a linear function of the applied magnetic field. Consequently, the magnetotropic susceptibility is also linear-in-field in this high-field regime. Figure~\ref{fig:fig2} shows that all field sweeps approach the same B-linear line at high magnetic fields, with smaller temperatures reaching it at smaller fields, and higher temperatures--at higher fields. 

We note that, despite being an idealized example, real spin-$\sfrac12$  systems might exhibit a behavior similar in magnitude and character to that shown in  Figure~\ref{fig:fig2} at temperatures and magnetic fields larger than the exchange energy scale in the system, hence the choice of the temperature range in Figure~\ref{fig:fig2}.

\subsection{ Magnetotropic susceptibility as a thermodynamic coefficient} 

Magnetotropic susceptibility is a thermodynamic coefficient, the second derivative of the free energy. As such, its behavior across the thermodynamic phase boundaries is constrained by general thermodynamic considerations. \cite{Callen1985} For example rotation of the sample in the applied magnetic field is described with a matrix of thermodynamic coefficients 
\begin{align}\label{eq:ehren}
	\bmatr{ dS \\ d\To_{\n} } 
	= \bmatr{ 
	\frac{C}{T} & ξ_{\n} \\ 
	ξ_{\n}         & k_{\n} 
	}
	\bmatr{ dT \\ dθ_{\n} } \,,
\end{align}
where $ξ_{\n} = dS/dθ_{\n} = dT_{\n}/dT$ is the rotational analog of magnetocaloric coefficients. 

Thermodynamic coefficients experience a discontinuous jump across the boundary of a continuous (second-order) phase transition. The magnitudes of the jumps in thermodynamic coefficients are related to one another via the Ehrenfest relations, which express the continuity of the thermodynamic potentials (first derivatives) across the boundary of a continuous phase transition. Equation~(\ref{eq:ehren}) relates the jump in the magnetotropic susceptibility to the jump in the heat capacity via $Δ k_{\n} = - Δ C/T_c × \para{dT_c/dθ_{\n}}^2$. Here $\para{dT_c/dθ_{\n}}$ is the change in the transition temperature when the sample is rotated in the applied magnetic field at a fixed temperature. The sign of the jump $Δ k_{\n}$ in the magnetotropic susceptibility across the phase boundary is fixed by thermodynamics. Similar to elastic moduli, or magnetic susceptibility, the magnetotropic susceptibility decreases as we enter into a lower symmetry phase.\cite{Modic2021}

\subsection{ Magnetotropic susceptibility as a thermodynamic response.} 
\label{sec:corrfunc-thermo}

In this section, we discuss the magnetotropic susceptibility in relation to microscopic degrees of freedom in the system, i.e., as a correlation function. This Section also sets up the discussion in the next section (Section~\ref{sec:corr-func-dynamic}), where we will introduce dynamics. 

The angular dependence of the free energy $F(∆θ)$ is generated by the term in the Hamiltonian $H_1(∆θ)$ that depends on the direction of magnetic field, 
	\begin{align}\label{eq:partition}
		e^{-β F(∆θ)} =& \Tr e^{ -β (H_0 + H_1(∆θ) ) } \,,
	\end{align} 
where $β=1/T$ is the inverse temperature (throughout this section the temperature is measured in energy units. In these units, $k_\Boltzmann = 1$). $\Tr ⋯ $ stands for the sum of expectation values over a complete set of states of the Hamiltonian, $\Tr ⋯ = ∑_n \bra{n}  ⋯ \ket{n} $. Here  $H_1(∆θ)$ is defined in such a way that $H_1(∆θ=0)=0$. 
The free energy can be expanded  in powers of $∆θ$ as 
\begin{align}
	F(∆θ) = F_0 + F_1  ∆θ + F_2 \; ∆θ^2/2\,. 
\end{align}
The magnetotropic susceptibility is equal to $F_2$. 

We want to express $F_2$ directly in terms of thermodynamic averages of the operators $H_0$ and $H_1(∆θ)$. Because operators $H_0$ and $H_1(∆θ)$ do not commute, the Taylor expansion of the matrix exponent in Eq.~(\ref{eq:partition}) in powers of $H_1(∆θ)$ has to be carried out by way of the identity $e^{A+B} = \lim_{n→∞} \para{e^{A/n} e^{B/n}}^n$ generated by the ``interaction representation''\cite{Negele1988}. Using the fact that $\exp\{-β F_0 \} = { \Tr \exp\{-β H_0\} }$, we can write  
	\begin{align}\label{eq:second-order}
		& e^{-β F_1∆θ -β(\sfrac{F_2}{2}) ∆θ^2 }   =  \brac{ \Tr e^{β H_0} } \brac{ \Tr e^{ -β( H_0 + H_1 ) } }  	 \notag\\
		& \qquad = 1+ 
				\av{\, -β H_1 \,} 
				+ \avav{\; (\!-β H_1\!),\,  (\!-β H_1\!) \; } \, + ⋯ \,,
	\end{align}
where $\av{ \,A\, } =  \Tr\!\!\brac{ \exp\{-β H_0\} A } \big/  \Tr\!\!\brac{\exp\{-β H_0\} }$ is the thermodynamic average and $\avav{A\;,\; B} $ is the average of the  (imaginary) time-ordered correlation function defined as 
	\begin{align} \label{eq:retardedf}
	\avav{A\;,\; B} =&  \; 2T^2\!\!\!∫\limits_0^{β}\!\!\!\!∫\limits_0^{β} \!\!\!dτ_1dτ_2\; θ(τ_1-τ_2) \; \av{ \,A(τ_1)B(τ_2)\,} \,,
	\end{align}
where $θ(τ_1-τ_2)= 1$  for  $τ_1>τ_2$ and $0$ otherwise and $A(τ)  =  \exp\{τ H_0\} A \exp\{-τ H_0\}$ is the (imaginary) ``time-evolution" of operator $A$ under the action of $H_0$. 
Because the thermodynamic average $\av{ \,A(τ_1)B(τ_2)\,}$ depends only on the difference $τ_1-τ_2$, the double integral in Eq.~(\ref{eq:retardedf}) can converted into a single integral~: 
\begin{align}
\avav{A\;,\; B} = 2T\!\!∫_0^{β}\!dτ \; (1-Tτ) \; \av{\,A(τ)B(0)\,} \,.
\end{align}
This equivalent form of $\avav{A\;,\; B}$ can be used in numeric calculations based on Eq.~(\ref{eq:second-order}). 

$F_1$ and $F_2$ in Eq.~(\ref{eq:second-order}) define the torque and magnetotropic susceptibility. To identify them, we exponentiate the second line in Eq.~(\ref{eq:second-order}) and keep terms that are linear and quadratic in $H_1$, 
\begin{align}\label{eq:exponentiate}
		& e^{-β F_1∆θ  -β(\sfrac{F_2}{2}) ∆θ^2}   \notag\\
		& \qquad = 
				e^{ \av{\, -β H_1 \,}  
				+ \avav{\; (\!-β H_1\!),\,  (\!-β H_1\!) \; } \, 
				- \sfrac{\av{\, -β H_1 \,}^2}{2} + ⋯ } \,,
\end{align}
where the last term in the second line is there to cancel the square of the first term in the expansion of the exponent. Magnetic torque and magnetotropic susceptibility can now be read off from Eq.~(\ref{eq:exponentiate}), 
	\begin{align}\label{eq:correlation-func}
		\To_{\n} =& \av{ \frac{dH}{dθ_{\n}} },\notag\\
			k_{\n} =& 
				\av{ \frac{d^2H}{dθ_{\n}^2} } 
					+\frac1T\av{ \frac{dH}{dθ_{\n}} }^2 
						- \frac1T \av{\!\!\av{  \frac{dH}{dθ_{\n}}  \,,\, \frac{dH}{dθ_{\n}}  }\!\!} \,.
	\end{align}
We have omitted the subscript in $H_1$  because $dH_1/dθ_{\n} = dH/dθ_{\n}$.

As an example, we consider a collection of spin-$\sfrac12$ spins with an anisotropic $g$-factor and exchange interactions, described by the Hamiltonian, 
	\begin{align}\label{eq:spin-hamiltonian}
H_0 = μ_{\Bohr}\B \!⋅\! \g \!⋅\! ∑_n (\bm{σ}^n/2) +  \frac14\!∑_{\av{n,m}} \bm{σ}^n 
\!⋅\! \hat{J}^{nm} \!⋅\!\bm{σ}^m\,,
	\end{align}
where $n,m$ represent different lattice sites and $\sfrac12 \bm{σ}^n$ is the spin operator on site $n$. Only the first ($g$-factor) term depends explicitly on the external magnetic field,  
	\begin{align}\label{eq:rotationofB}
H_1(θ\n) = &  - (δ_{θ\n}\B) ⋅ \Mop \,, \quad \Mop = -μ_{\Bohr} \; \g  ⋅  ∑_n \bm{σ}^n/2 \,,
	\end{align}
where $ \Mop $ is the magnetization operator. We need to expand $δ_{θ\n}\B$ to second order in the rotation angle $∆θ$,  
	\begin{align}\label{eq:justrotationofB}
δ_{θ\n}\B =&  ∆θ (i\N)⋅\B + \frac{∆θ^2}2  (i\N)⋅(i\N)⋅\B\,. 
	\end{align}
We have
	\begin{align} \label{eq:torque-op}
	& \frac{dH}{dθ} =  \Top_μ n_μ=   - \B \;⋅  (i\S_μ n_μ)  ⋅ \Mop 
				=  (\Mop × \B)_μ ⋅ n_μ \,,
	\end{align}
where $\Top_μ$ is the torque operator, and 
	\begin{align} \label{eq:quadraticH}
\frac{d^2H}{dθ^2} = &  -  \, \B ⋅  (i\S_μ n_μ)⋅  (i\S_ν n_ν) ⋅  \Mop   	\,.		 
	\end{align}
Substituting these into  Eq.~(\ref{eq:correlation-func}) we obtain 
	\begin{align} \label{eq:other3}
		\To_{μ} =& \av{\, \Top_μ  \, } = - \B \;⋅  (i\S_μ)  ⋅ \av{\, \Mop\,} \,, \\
   		  \label{eq:other2}
   		  k_{μν} =& \frac12   \B ⋅ (\S_μ ⋅ \S_ν + \S_ν ⋅ \S_μ) ⋅ \av{\,\Mop\, } 
   		  \notag\\
					&\qquad + \frac1T \av{\,  \Top_μ \, } \av{\, \Top_ν \,}    
					 - \frac1T \av{\av{\, \Top_μ \;,\; \Top_{ν} \, } }   \,.
	\end{align}
The first line in $k_{μν} $ is equal to the first line in Eq.~(\ref{eq:definition1}).  
Because the torque operator is $\Top = - \Mop  ⋅  (\n  ×  \B)$, Eq.~(\ref{eq:torque-op}), the second line has the same structure as the microscopic correlation function for magnetic susceptibility, 
	\begin{align}\label{eq:chidef}
		{χ}_{μν}  = 	 
	\frac1T \av{\Mop_{μ} } \av{\Mop_{ν}}
	-	\frac1T \avav{ \Mop_{μ} \;,\; \Mop_{ν}  }   \,.
	\end{align}
Substituting Eq.~(\ref{eq:torque-op}) into Eq.~(\ref{eq:other2}) and using Eq.~(\ref{eq:chidef}) one can see that the second line in Eq.~(\ref{eq:other2}) is equal to  $-(  \n×\B) ⋅ \hat{χ} ⋅ (\n×\B) $, the second term in Eq.~(\ref{eq:fromfree1}). 

Equation~(\ref{eq:other2}) applies more generally, beyond the specific example it was derived for. The magnetotropic susceptibility has two qualitatively different parts as will become clear in the discussion of dynamic magnetotropic susceptibility in Section \ref{sec:corr-func-dynamic}. The linear-in-magnetization part in Eq.~(\ref{eq:other2}) originates from the fact that the magnetic torque operator depends explicitly on the external magnetic field, described by the second derivative term $d^2H/dθ^2$ in Eqs.~(\ref{eq:torque-op}), (\ref{eq:correlation-func}), and (\ref{eq:quadraticH}). It does not describe the actual response of the system to rotation in the applied field. Instead, it captures the ``redefinition’’ of the torque operator in the rotated reference frame. 
The second term in Eq.~(\ref{eq:other2}) describes the proper response function part of the magnetotropic susceptibility given by the torque-torque correlation function, as required by fluctuation-dissipation analysis \cite{Onsager1930, Onsager1931}.

\subsection{Rotation-in-field-induced relaxation phenomena captured in the frequency-dependent magnetotropic susceptibility.}
\label{sec:corr-func-dynamic}

Given the nature of the experimental setup in which magnetotropic susceptibility arises (Section \ref{sec:frequencyshiftmeas}), we now discuss the dynamic---frequency-dependent---magnetotropic susceptibility $k_{\n}(\B,ω)$. It captures the relaxation phenomena coupled to the rotation of the sample in an applied magnetic field. 
The time-dispersed magnetotropic susceptibility $k_{\n}(t-t')$ describes time-delayed response of the torque $Δ\To_{\n}(\B)$ to rotation of the sample in applied magnetic field, 
\begin{align}\label{eq:dynamick}
	Δ\To_{\n}(\B,t) = ∫\limits_{-∞}^{t} \!\!\!dt' \; k_{\n}(\B, t-t') \; Δθ_{\n}(t')  \,.
\end{align} 
Its Fourier transform defines the dynamic magnetotropic susceptibility $k_{\n}(\B,ω) = ∫_{0}^{∞}dt \exp({iω t}) k_{\n}(\B, t)$. 

Kubo analysis \cite{Negele1988}  starting from Eq.~(\ref{eq:dynamick}) suggests that the dynamic response $k_{\n}(ω) $ is equal to the dynamic torque-torque correlation function,  $k_{\n}(ω) =  ∫_{0}^{∞}dt \exp({iω t})  \av{ \Top_{\n}(t) \Top_{\n}(0) } $. This is because the torque operator is conjugate to the angle variable $Δθ_{\n}$ in the microscopic Hamiltonian, Eq.~(\ref{eq:torque-op}), and therefore, determines the time-evolution of the torque under time-dependent $Δθ(t)$. This is similar to the microscopic interpretation of the dynamic magnetic susceptibility as a correlation function of magnetization operators. 
This argument captures the entire imaginary part of the dynamic magnetotropic susceptibility $k_{\n}(\B,ω)$ as well as that part of the real part of $k_{\n}(\B,ω)$ that is related to the imaginary part by Kramers-Kronig (analyticity) requirements. In particular, it shows that the dynamic magnetotropic susceptibility $k_{\n}(ω)$  is identical in its analytic properties to the dynamic magnetic susceptibility $χ(ω)$ and contains the same physical information. Direct calculation of the dynamic torque-torque correlation  function results in 
\begin{align}\label{eq:imaginaryk}
\Im k_{\n}(ω) = - (\n×\B) ⋅ \Im\hat{χ}(ω) ⋅ (\n×\B) \,.	
\end{align}

However, without $\Im$, Eq.~(\ref{eq:imaginaryk}) does not produce a correct relation between $k_{\n}$ and  $\hat{χ}$ in the static limit $ω=0$ given by Eq.~(\ref{eq:fromfree1}). This is because the time-dispersed magnetotropic susceptibility $k_{\n}(t-t')$ contains an instantaneous part, proportional to delta-function $δ(t-t')$ associated with the direct dependence of the magnetic torque on magnetic field, Eq.~(\ref{eq:torque-def}).  

The instantaneous part in $k_{\n}(t-t')$ produces a frequency-independent real function on the real axis of frequency, in $k_{\n}(ω)$. Such a constant function has zero imaginary part everywhere in the complex plane. 
The instantaneous term in the magnetic torque originates in the explicit dependence of magnetic torque operator $\Top_{\n}(\B) =  -\Mop ⋅  (\n ×\B )$ on applied magnetic field $\B$. This term is reactive---it is not associated with the time evolution of torque under the action of time-dependent Hamiltonian, as captured by the Kubo argument. The complete  relation between the dynamic magnetic susceptibility $\hat{χ}$ and magnetotropic susceptibility $k_{\n}$ is given by 
		\begin{align}\label{eq:fromfree-dynamic}
&	k_{\n}(\B, ω) \!=\!  (\n×\B)  ⋅  (\n×\M)  \notag\\
&	\qquad\qquad\qquad 	- (\n×\B)  ⋅ \hat{χ}(\B,ω) ⋅ (\n×\B)	\,.
		\end{align}
The first term here is not associated with relaxation- or dispersion phenomena in the sample. It is a mere redefinition of the torque operator in the rotated frame of reference. The second term has the structure of dynamic correlation function\cite{Negele1988} of magnetic torque operators, $\Top_{\n}(\B) =  -\Mop ⋅  (\n ×\B )$. 


%

\section{The magnetotropic susceptibility in frequency-shift measurements.} 
\label{sec:frequencyshiftmeas}

\subsection{The frequency shift of an oscillating cantilever-sample assembly.}
\label{sec:frequencyshift}

The magnetotropic susceptibility emerges naturally in frequency shift measurements of oscillating cantilever-sample assembly in a uniform applied magnetic field  \cite{Bishop1978, Kleiman1984, Kleiman1985, Kleiman1987, Worthington1987, Bishop1992, Bolle1999, Chiaverini2001,  Modic2018a, Modic2021, Pocs2021, Mumford2021, Maeno2011}. In these measurements, the sample is attached at the free end of a cantilever. The oscillation amplitude is read out by a piezo-mechanical transducer \cite{Bishop1978, Kleiman1984, Kleiman1985, Kleiman1987, Worthington1987, Bishop1992, Bolle1999, Chiaverini2001,  Modic2018a, Modic2021, Pocs2021} or by laser optics \cite{Mumford2021,Maeno2011}. When driven near its $n$th mechanical resonance mode at frequency $ω_{n}$, the cantilever behaves as a simple harmonic oscillator with mechanical energy 
	\begin{align}\label{eq:effective0}
  E= \frac{I_n}2 \left(\frac{dΔθ}{dt}\right)^2 + \frac{\St_n}2 Δθ^2  \,,
	\end{align}
where $Δθ$ is the bending angle at the free end of the cantilever. This defines the effective bending stiffness $\St_n$ and the effective moment of inertia $I_n$ of a cantilever near its $n$-th resonance mode. $\St_n$ and $I_n$ depend on the resonance mode $n$ as well as on the geometry and the construction of the cantilever (Section~\ref{sec:cantilever}). The resonance frequency $ω_{n}$ of the cantilever is determined by the bending stiffness $\St_n$ and the effective moment of inertia $I_n$ as $ω_{n} = \; \sqrt{\fra{ \St_{n} } {I_{n}}} $. 

For a thin cantilever driven near the fundamental ``flapping'’ mode  (Section~\ref{sec:cantilever},  Figure \ref{fig:fig3}), 
\begin{align}\label{eq:flapping}
& \St_{0} \;= 1.63 \; \frac{γ}{L} = 0.136 \; \frac{wh^3Y}{L} \,, \qquad 
I_{0} = 0.132 \;ρ whL^3\,, \quad \notag\\
& \qquad 
ω_{0} = \; \sqrt{\frac{ \St_{0} } {I_{0}}} \;=\; 1.02 \; \frac{h}{L^2}  \, \sqrt{\frac{Y}{ρ}}\,,
\end{align}
where $Y$ is the Young's modulus of the cantilever, $ρ$ is its density, $w$ is the width, $L$ is its length, and $h$ is its thickness. $γ$ is the arc stiffness of the cantilever discussed in Section \ref{sec:cantilever}. 

In the experiments where the displacement of the cantilever is detected optically \cite{Maeno2011, Mumford2021}, it might be convenient to define the energy of the cantilever in terms of the displacement $Δ X$ at the tip of the cantilever rather than the rotation angle $Δθ$,
\begin{align}\label{eq:EofX}
		E= {\frac{m_n}2} \para{\frac{dΔ X}{dt}}^2 + {\frac{c_n}{2}} {Δ X^2} \,,
\end{align}
where $m_n$ is the effective $n$-th mode mass  coefficient, and $c_n$ is the effective $n$-th mode spring constant. For an oscillating cantilever, $Δ X$ and  $Δθ$ are proportional to each other, with a mode-dependent coefficient (Figure \ref{fig:fig3}). For the fundamental mode~:
\begin{align}\label{eq:deltaX}
	Δ X = 0.73 L\, Δθ\,. 
\end{align}
For higher-$n$ modes, the coefficient of proportionality is given in Eq.~(\ref{eq:effectivelength}). The spring constant of the fundamental mode, $c_0=\St_0 / (0.73 L)^2 = 3.06 \; γ /L^3 = 0.255 \; Y w h^3/L^3$,  \cite{Maeno2011}  is obtained by comparing Eqs.~(\ref{eq:effective0}) and (\ref{eq:EofX}). The effective mass coefficient $m_0$ of the fundamental mode in Eq.~(\ref{eq:EofX}) is given by $m_0 = I_0/(0.73 L)^2 = 0.248 \; m_{\Lever}$ where $m_{\Lever}$ is the mass of the cantilever.  In Section \ref{sec:static-calibration} we discuss calibrating the arc stiffness $γ$ of the cantilever in static measurements.  The bending stiffness  $\St_n$ and spring constant $c_n$ are then determined by the arc stiffness $γ$.

When a sample is attached at the tip of the cantilever, the angular dependence of the energy of the cantilever-sample assembly acquires a small additional magnetic field-dependent potential energy, originating from the magnetically-anisotropic free energy of the sample, 
	\begin{align}\label{eq:sample-assembly}
δ E_{\text{sample}} =  \To(\B) \;Δθ + k(\B) \;\frac{Δθ^2}2 \,.
	\end{align}
The effect of the $Δθ$-linear term is to shift the equilibrium value of $Δθ$ away from zero, by an amount proportional to the torque $\To(\B)$ on the sample. The effect of the $Δθ$-bilinear term is to shift the frequency of the mechanical resonance of the cantilever, determined now by the combined effect of the bending stiffness $\St_n$ of the cantilever and the magnetotropic susceptibility $k(\B)$ of the sample---both describe the quadratic-in-$Δθ$ change in energy in  Eq.~(\ref{eq:effective0}). For arbitrary magnitude of the magnetotropic susceptibility of the sample, the frequency shift $Δω$ is determined by 
\begin{align}\label{eq:nonlinear-freq-shift}
(ω_0+Δω)^2 = \frac{k(\B) + \St_n}{I_n} \,. 	
\end{align}
When the magnetotropic susceptibility of the sample is small compared to the bending stiffness of the cantilever, $k(\B) ≪ \St_n$, the frequency shift $Δω$ is small compared to $ω_0$, and we can expand,
	\begin{align}\label{eq:freqshift}
\frac{Δω}{ω_0} ≈ \frac{k(\B)}{2\St_n}\,, \qquad   k(\B)≪ \St_n \,.
	\end{align}
The magnetotropic susceptibility of the sample is proportional to the shift of the resonance frequency of the cantilever-sample assembly in magnetic field. 

Equation~(\ref{eq:nonlinear-freq-shift})  assumes that the mass of the sample, $m_{\Sample}$, is much smaller than the mass of the cantilever, $m_{\Lever}$. When considering the finite mass of the sample, an additional kinetic energy term is introduced in Eqs.~(\ref{eq:effective0}) and (\ref{eq:sample-assembly}), expressed as $m_{\Sample}/2 \times  (dΔ X/dt)^2$. This term is equivalent to an increase in the effective moment of inertia, $I_n$. For instance, for the fundamental mode, $I_0$ becomes $I_0 + m_{\Sample} (0.73 L)^2$. It is important to note that when this change in the effective moment of inertia is small, it does not affect the relative frequency shift in Eq.~(\ref{eq:freqshift}). 

\subsection{Examples}
\label{sec:examples}

To illustrate the quantitative aspects of the resonance frequency shift measurements, we consider two studies: an interacting system of spin-$\sfrac12$'s in RuCl3 \cite{Modic2021} and quantum fluxes in a superconducting ring \cite{Maeno2011}.

The silicon cantilever in Ref.~\onlinecite{Modic2021} measures 3.4 $μ$m in thickness, 60 $μ$m in width, and 300 $μ$m in length. From Eq.~(\ref{eq:flapping}), the bending stiffness is 180~nJ and the frequency of the fundamental resonance mode is close to 50~kHz. At cryogenic temperatures in low-pressure exchange gas, the cantilever has a $Q$-factor of around 3 $× 10^4$ (Section~\ref{sec:friction}) and a resonance width of a few hertz. 

Below 5 K,  RuCl$_3$ is characterized by an anisotropic magnetic susceptibility, $χ_a-χ_c=$ of 0.06 J/mol-f.u.T$^2$, or 0.01 $μ_{\Bohr}$/T per Ru spin. The RuCl$_3$ sample in Ref.~\onlinecite{Modic2021}  measures 50 $×$ 70 $×$ 2 $μ$m$^3$ with a mass of 20 nanogram or 0.1 nanomol-f.u. molar mass, has magnetic susceptibility of 6 pJ/T$^2$. In 1~T  magnetic field along the $a$ axis, the magnetotropic susceptibility of the sample is 6~pJ, from Eq.~(\ref{eq:shminear}). This shifts the resonance frequency of the 180~nJ-stiffness cantilever by 0.8~Hz (Eq.~\ref{eq:freqshift}), comparable with the resonance width. Such frequency shifts are well within detection limits. The frequency shift changes sign to  -0.8~Hz, when the same magnetic field is applied along the $c$-axis, as long as we stay in the linear regime at this field along the $c$-axis. The frequency shift increases quadratically with the magnetic field, Eq.~(\ref{eq:shminear}), as long as we stay in the linear regime. 

At even higher magnetic fields, beyond the energy of the effective spin-exchange interaction and the thermal energy, $k_{\Boltz} T$, the spin-$\sfrac12$ system in RuCl$_3$ becomes scale-invariant, i.e., the free energy is a function of external energy scales only. This implies that the free energy and magnetotropic susceptibility are linear function of magnetic field at low temperatures and large magnetic fields (Section \ref{sec:spin-one-half}) \cite{Modic2021}.

A 0.1 nanomol-f.u. system of non-interacting spin-½’s with a $g$-factor anisotropy of 2 (Figure~\ref{fig:fig2}) has the magnetotropic susceptibility of 10~nJ at 10 K and 25 T, and at 50 K and 50 T. For the cantilever of 180~nJ bending stiffness, the magnetotropic susceptibility of 10 nJ would result in a frequency shift close to 3 kHz. Frequency shifts of similar magnitude were  observed in RuCl$_3$ sample of similar molar mass \cite{Modic2021}. 

The finite mass of the cantilever, 150~ng, results in a maximum gravitational frequency shift of 10~mHz, Eq.~\ref{eq:gravitational-shift-cantilever}, small compared to the frequency shift associated with magnetic anisotropy of the sample. 

The second example is the study of quantum jumps of magnetic flux in a small superconducting ring, reported in Ref.~\onlinecite{Maeno2011}. In this paper, a much thinner silicon cantilever was employed, with dimensions of 0.1~$μ$m thickness, 3~$μ$m width, and 80 $μ$m length. The bending stiffness of such cantilever is 1~pJ and the resonance frequency 20~kHz, as determined by Eq.~\ref{eq:flapping}. The observed resonance frequency is somewhat lower at 16~kHz, possibly due to the added inertia of the sample mass of 50~pg. The spring constant measured in Ref.~\onlinecite{Maeno2011}, is 3.6 $× 10^{-4}$ N/m, which corresponds to an effective bending stiffness of 1.2~pJ for the fundamental mode, close to our estimate (see Section~\ref{sec:static-calibration} for the relation between the two). The observed $Q$-factor of 6.5~$× 10^4$ corresponds to a resonance width of 0.3~Hz. 

We can estimate the energy change that accompanies a single quantum flux entering the superconducting ring as $Δ_{\flux} E ≃ - Φ_0 Φ_{\ext} / d $ \cite{Tinkham1996}. Here, $Φ_0$ is the flux quantum, $Φ_{\ext}$ is the flux of the external magnetic field crossing the superconducting ring, and $d$ is the diameter of the ring. The angular dependence of $Δ_{\flux}E(θ)$, is the same as the angular dependence of $Φ_{\ext}$ at a fixed external magnetic field, proportional to $\cosθ$. Between the jumps, the magneto-static energy of the superconducting ring changes quadratically with field with a coefficient proportional to $\cos2θ$. This estimate does not account for the impact of a finite (and anisotropic) penetration depth or the finite difference between the ring's inner and outer diameters. We note that $Δ_{\flux} E$ is negative, in accordance with Le Chatelier's principle. This requires that the frequency must jump up as one extra flux enters the ring, consistent with the measurements \cite{Maeno2011}. 

Given the 1~$μ$m diameter of the superconducting ring, the jump in magnetotropic susceptibility, calculated using Eq.~(\ref{eq:fromfree1}), $Δ k = d^2 Δ_{\flux}E(θ) / dθ^2$, is $Δ_{\flux}k = + Φ_0 Φ_{\ext} /d = 1 × 10^{-4}$ pJ when the external field component perpendicular to the loop is 10~G. For a cantilever of bending stiffness  1~pJ  and a resonance frequency of 20~kHz, such a jump in the magnetotropic susceptibility of the superconducting ring corresponds to a frequency shift of 1~Hz, comparable to the resonance width of 0.3 Hz \cite{Maeno2011}. If, instead, we used the thicker cantilever from the first example, the frequency shift corresponding to the flux jump would be 1.5 $× 10^{-5}$~Hz. 

The mass of the cantilever is 50~pg, which results in a maximum gravitational frequency shift of 0.1 Hz (Eq.~\ref{eq:gravitational-shift-cantilever}), comparable to the resonance shift in the flux-jump. Consequently, in this measurement, the angular dependence of the frequency shift has a smooth, field-independent background of a magnitude comparable to that of the flux-jump shifts, both having the same angular dependence $\propto\cosθ$. The gravitational shift, associated with the sample mass of roughly 50~pg as in Ref.~\onlinecite{Maeno2011}, has a similar magnitude and the same angular dependence (Eq.~\ref{eq:gravitational-shift-sample}).

At cryogenic temperatures in vacuum, the resonance width is determined by the thermoelastic friction in the cantilever (Section~\ref{sec:friction}) and the energy dissipation within the superconducting ring. We can estimate the thermoelastic friction in the cantilever using Eqs.~(\ref{eq:functionA}) and (\ref{eq:thermoelasticQ}). The heat diffusion time $τ_h$ across the cantilever's thickness $h$ is $3(h/c)^2/τ_e$, where $τ_e$ is the mean free time of phonons in silicon at cryogenic temperatures. Assuming  $τ_h$ of about 10 ps, the resonance frequency of 20 kHz, and a thermodynamic factor in Eq.~(\ref{eq:thermoelasticQ})  of 1\% \cite{Callen1985}, the estimated thermoelastic friction limit on the $Q$-factor is around $10^9$. For the thicker lever in Ref.~\onlinecite{Modic2021}, the thermoelastic friction puts a stricter limit on the $Q$-factor, of a few times $10^6$. The observed $Q$-factor of 6.5 $× 10^4$ suggests that the thermoelastic friction is not the limiting factor on the resonance width. The observed resonance width of 0.3 Hz might originate from a small viscous friction in the surrounding exchange gas, in which case, the resonance width would be independent of magnetic field. 

Alternatively, the resonance width could originate from the energy dissipation in the superconducting ring oscillating in an external magnetic field, via the imaginary part of the magnetotropic susceptibility of the superconducting ring in Eq.~(\ref{eq:resonace-width-intrinsic}). In this scenario, the resonance width will depend on an applied magnetic field.  If the resonance width $ΔΓ$ originates in the dissipative phenomena in the superconducting ring, an accompanying shift in resonance, $Δf$, of comparable magnitude is expected, i.e., $Δf ≃ ΔΓ$. This is because the measured response $f(ω)$ in Eq.~(\ref{eq:resonace-width-intrinsic}) is an analytic function limited by causality (Kramers-Kronig relations). It would be interesting to see how much of the observed frequency shift is associated with the dissipation in the superconducting ring associated with the flux jump. 

\subsection{Arc stiffness of the cantilever in the static measurements. }
\label{sec:static-calibration}

The effective bending stiffness $\St_0$, or the effective spring constant $c_0$, can be inferred from the frequency shift in uniform gravity or uniform magnetic field gradient (Section \ref{sec:gravity}). They can also be calibrated by applying a static force or static torque at the tip of the cantilever, or by using the weight of the cantilever as a source of force. In each of these measurements, one can determine arc stiffness $γ$ of the cantilever and, through it, the effective bending stiffness $\St_0$, Eq.~(\ref{eq:effectiveK}) or the spring constant $c_0$, Eq.~(\ref{eq:EofX}).

Applying a static torque $T$ at the free end of the cantilever creates a uniform cross-sectional torque, Eq.~(\ref{eq:crossectional-torque})  (Figure~(\ref{fig:fig3}) along the cantilever, $T(z)=T$, and, therefore, uniform curvature $\lap ζ(z)$. The bending shape of the cantilever is a parabola $ζ(z) = T/γ  ×  z^2/2$ for which $ΔX = (1/2) L\, Δθ$.  The bending angle at the free end of the cantilever is $Δθ = 2ΔX/L  = T L / γ$. 

Applying a static force $F$ along the $x$-axis (Figure~(\ref{fig:fig3}) at the free end of the cantilever, creates the linear-in-$z$ cross-sectional torque, $T(z) = F(L-z)$. The bending shape of the cantilever is a cubic parabola, $ζ(z) = FL/γ \,×\, (3-z/L)\,  z^2/6$, for which $ΔX = (2/3) L Δθ$. The displacement at the free end of the cantilever is  $ΔX = 1/3 \,×\,  FL^3/γ$, and, therefore, the spring constant $c=F/ΔX$ is $c=3γ/L^3$. 

If the weight of the cantilever in Earth's gravity is the source of static force, the cross-sectional torque is $T(z) = m_{\Lever}g \cosθ \,×\, (L-z)^2/2L$, where $θ$ is the angle between the cantilever and the direction of the force of gravity. The  displacement is $ζ(z) = m_{\Lever}gL^3/γ \,×\, \cosθ \,×\, [\;6\; (\!z\!/\!L\!)^2 \,-\, 4 \, (\!z\!/\!L\!)^3 \,+\, (\!z\!/\!L\!)^4 \,]\,/\,24$, with $ΔX = (3/4) L \, Δθ$. The displacement at the free end of the cantilever is $ΔX = 1/8 \,×\,  m_{\Lever}gL^3 / γ \,×\, \cosθ $.

\subsection{Effects of friction in the sample and in the cantilever and the resonance width.}
\label{sec:friction}

The width of the resonance is determined by those relaxation phenomena in the sample and the cantilever that are coupled effectively to the rotation of the sample and the cantilever in the applied magnetic field. It is important for this discussion that the measurement of the mechanical oscillations of the cantilever produces a response function with the same analytic properties as the magnetotropic susceptibility of the sample.  

For the sake of this discussion, it suffices to model the amplitude readout in these experiments to be proportional to the rotation angle $Δθ(t)$ at the tip of the cantilever.  For example, in optical-readout setups \cite{Maeno2011, Mumford2021} the readout is proportional to the displacement $Δ X(t)$ at the free end of the cantilever. In piezo-readout setups, (e.g., Ref.~\onlinecite{Modic2021}), the measured signal is proportional to the stress induced by the lever at the attachment end of the cantilever, which, in turn, is proportional to the cross-sectional torque $T(z=0)$ at the attached end of the cantilever (see Eq.~(\ref{eq:crossectional-torque}) for details).  All these are proportional to each other for small displacements of the cantilever.

Therefore, quite generally, the measured response is proportional to the mechanical response $f(ω) =  ∫_{0}^{∞} d(t-t')\, \exp({iω(t-t')}) Δθ(t)\,Δθ(t') $. Near the mechanical resonance, Eqs.~(\ref{eq:effective0}) and (\ref{eq:freqshift}), $ω_n^2 = [\St_n(ω) + k(ω)]/I_n$ the response function $f(ω)$ has a broadened Lorentzian form 
\begin{align}\label{eq:piezocircuitS}
	f(ω) ∝ & \frac1{ ω^2 - [\St_n(ω) + k(ω)]/I_n }  \notag\\
		= & \frac1{ ω^2 - ω_0^2+ i ω_0Γ } 	\,.
\end{align}  
The relaxation phenomena in the sample and in the cantilever shift the poles of the response function $f(ω)$ of the cantilever below the real axis, $± ω_0 - i Γ/2$, where $Γ$ is the width of the resonance.
The width of the resonance, $Γ$, is determined by the imaginary part of $ω_0^2 = [\St_0(ω) + k(ω)]/I_0$. For the fundamental mode, 
\begin{align}\label{eq:resonace-width-intrinsic}
Γ = - \frac{1}{I_0 ω_0} \brac{ \Im \St_0(ω_0) + \Im k(ω_0) }  \,. 	
\end{align}
The partial resonance width produced by relaxation phenomena in the sample is proportional to the imaginary part of the dynamic magnetotropic susceptibility $k(ω_0)$ discussed in Section~\ref{sec:corr-func-dynamic}. 

We note that Eq.~(\ref{eq:piezocircuitS}) implies that relaxation phenomena in the sample result not only in resonance width broadening proportional to the imaginary part of magnetotropic susceptibility $\Im k(ω_0)$ but also in a finite frequency shift $Δω ∝\Re k(ω_0)$ associated with $\Im k(ω_0)$ via Kramers-Kronig (analyticity) relations. These considerations become especially important in experiments where the frequency shifts are exceedingly small. \cite{Maeno2011} 

Equation~(\ref{eq:piezocircuitS}) can also be used as a starting point for discussion of the frequency shift and resonance width broadening in a polycrystal.  The "thermodynamic" part of the magnetotropic susceptibility vanishes in a polycrystal. The magnetotropic susceptibility in the polycrystal is determined entirely by the relaxation phenomena in the sample. For a polycrystal, the entire frequency shift in  Eq.~(\ref{eq:piezocircuitS})  is related to the resonance width broadening by Kramers-Kronig relations.\cite{Negele1988}
    
We now discuss the relaxation phenomena in the cantilever captured in the imaginary part of the dynamic bending stiffness $\St(ω)$. In complete analogy to the discussion in Section \ref{sec:corr-func-dynamic}, in the presence of slow relaxation phenomena in the cantilever, its mechanical response of the cantilever (e.g., stress) is no longer an instantaneous function of the bending angle $Δθ(t)$. In particular, one cannot, strictly speaking, define an instantaneous effective energy Eq.~(\ref{eq:effective0}). 
This situation is described by introducing a variable $ΔΘ = dE/dΔθ$, which is thermodynamically conjugate to the bending angle $Δθ$ at the free end of the cantilever.  Equation~(\ref{eq:effective0}), defines the instantaneous response of $Θ$ to the bending angle $Δθ$ in the absence of relaxation phenomena, $Θ(t) = \St Δθ(t)$. Dynamic bending stiffness $\St(ω)$ defines the time-dispersed, non-instantaneous, relation between $ΔΘ(t)$ and $Δθ(t')$~:
\begin{align}
ΔΘ(t) = ∫_{-∞}^{t} \!\!\!dt' \; \St(t-t') \; Δθ(t') \,. 	
\end{align}
Dynamic bending stiffness $\St(ω)$ is the Fourier transform of its time-dispersed counterpart, 
$\St(ω)=  ∫_{0}^{∞}dt \exp({iω t})  \St(t) $\,. 	

Relaxation phenomena in the cantilever include thermoelastic friction\cite{Zener-I-1937, Zener-II-1937} and the viscous friction in the surrounding exchange gas.\cite{Chadwick2008, Maali2005}  
The motion of the cantilever in the surrounding He gas or liquid leads to viscous friction. For a 100 x 50 $μ$m$^2$ cantilever oscillating at 10 kHz, the Reynold's number is small, $∼0.5$ in air and $∼5$ in liquid He-4 at 2K.\cite{Keesom1938}  Therefore, we can use the Stokes formula for the viscous force. The ratio of the resonance width to the resonance frequency can be estimated as the ratio of the energy dissipated per cycle to the energy stored, $Γ/f =  \para{ 6πη R f u^2}/\para{ \St_0 (u/L)^2 } = { 6πη  \sqrt{wL} f L^2}/\St$ where $u$ is the amplitude of oscillations. For a 300 x 60 $μ$m$^2$ cantilever the viscous resonance width is about $30$ Hz in the air at room temperature and atmospheric pressure and about $3$ Hz in liquid He-4 at 2K.\cite{Keesom1938} 

Regardless, the $Q$-factor of the cantilever is intrinsically limited by internal thermoelastic friction.\cite{Zener-I-1937, Zener-II-1937}  Bending of the cantilever induces nonuniform strain across the thickness of the cantilever, which, via thermal expansion coefficients, induces temperature gradient across the thickness of the cantilever,  $δ T ∝ x$. Irreversible relaxation of the temperature gradient via heat diffusion across the thickness of the cantilever results in entropy production and, therefore, energy dissipation. The effect of such thermoelastic friction on the dynamic bending stiffness $\St(ω)$ can be captured  (via Eq.~(\ref{eq:effectiveK})) by the dynamic (frequency dependent) Young modulus, $Y(ω) = Y_T + (Y_S - Y_T) A(ω)$, where $Y_S$ and $Y_T$ are adiabatic and isothermal values of Young's modulus respectively and $A(ω)$ is a dimensionless factor normalized as $A(ω=0)=0$ and $A(ω=∞) =1$. The value of ${\fra{(Y_S-Y_T)}{Y_T}}$ is about 1 \% at room temperature and smaller at lower temperatures.\cite{Callen1985} For a thin cantilever, 
\begin{align}\label{eq:functionA}
A(ω) =&  48\!\!\!\!\!\!\! ∑_{k_n=(2n+1)π}  \frac{1}{k_n^{4}} \; \frac{-iω τ_h}{ -iω τ_h  + k_n^2 }   \notag\\
= & 1-\frac{12}{-iω τ_h} + \frac{24}{(\sqrt{-iω τ_h})^{3}}\tanh\frac{\sqrt{-iω τ_h}}{2} \,,
\end{align}
where $τ_h$ is the heat diffusion time across the thickness of the cantilever, $1/τ_h = D/h^2$. Here $D = κ/C_V$ is the heat diffusion coefficient. The $Q$-factor~:
\begin{align}\label{eq:thermoelasticQ}
Q^{-1} = -\frac{Y_S-Y_T}{Y_T} \; \Im{A(ω)}\,.
\end{align} 

We can estimate $τ_h$ as $3 (h/c)^2 / τ_e$ where   $τ_e$ is the mean free time of phonons in the cantilever and $h/c$ is the time it takes for sound to traverse the thickness of the cantilever. For thin cantilevers, $ω τ_h$ is small and Eq.~(\ref{eq:functionA}) behaves as $A(ω) = -i ω τ_h /10 $. The $Q$-factor of a 1-micron-thick silicon cantilever with a resonant frequency of 100 kHz and phonon mean free path of 10 microns at cryogenic temperatures is limited by $10^7$. 

\subsection{Adiabatic evolution of freely oscillating cantilever.}

The frequency shift can be detected using lock-in techniques.\cite{Bishop1992, Kleiman1985, Modic2018a, Modic2021,Mumford2021, Maeno2011}  In some experiments, such as in pulsed magnetic fields, it is more straightforward to observe the free oscillation of the cantilever in the time-dependent magnetic field. \cite{Modic2021} Here we discuss some general properties of the oscillating cantilever relevant to such measurements. The free evolution of the cantilever in a slowly changing environment is described by 
	\begin{align}\label{eq:timedependent} 
 \frac{d^2∆θ}{dt^2} +ω_0(t)^2∆θ =0 \,, 
	\end{align}
where the time dependence of the ``spring constant" parameter, 
\begin{align} 
	ω_0(t)^2 = [{\St+k(t)}]/{I}\,, 
\end{align} 
is determined by the slow evolution of the magnetotropic susceptibility $k(t) = k(B(t)$ in an external magnetic field $B(t)$. When the magnetic field changes slowly, the oscillation frequency of the cantilever adiabaticallyfollows  the instantaneous value of $ω_0(t)$ --- the instantaneously observed oscillation frequency is equal to the parameter $ω_0(t)$ in Eq.~(\ref{eq:timedependent})~:
	\begin{align}
∆θ(t) =& A(t) e^{ iφ(t) }   \,, \quad  φ(t) = ∫\limits^tω_0(t')dt'\,.
	\end{align}
The quantitative measure of the slowness of $ω_0(t)$ is the adiabaticity parameter $α = d(ω_0^{-1})/dt$ defined as the fractional change of frequency in one oscillation period.\cite{Landau-Mechanics1976} 

While the observed frequency of the cantilever follows adiabatically the instantaneous value $ω_0(t)$ determined by the magnetic field $B(t)$ and the magnetic response of the sample, the amplitude $A(t)$ of the oscillation decays. In a typical setup, the source of amplitude decay is energy dissipation in the surrounding exchange gas, in the components of the setup susceptible to irreversible deformations, as well as, and possibly dominantly so at low temperatures, energy dissipation associated with thermoelastic friction and solid viscosity in the cantilever itself (Section \ref{sec:friction}). 

The amplitude of oscillations of the cantilever evolves reversibly (as well as irreversibly, due to energy dissipation) in a time-dependent magnetic field. This is because the cantilever-sample assembly in an external magnetic field is not a closed mechanical system. The mechanical energy of the sample-cantilever assembly changes by the amount of work done by the magnetic field on the sample. In the adiabatic regime, this reversible energy exchange is captured by the weak time dependence of the adiabatic invariant,
\begin{align}
J=\frac{E(t)}{ω_0(t)}	\,,
\end{align}
 which does not change when the spring constant parameter $ω_0(t)^2$ varies slowly in time \cite{Landau-Mechanics1976}.  Here $E(t)=I\,ω_0(t)^2A(t)^2$ is the average mechanical energy of the cantilever. Therefore, the oscillation amplitude of the cantilever will change reversibly according to
 \begin{align}
 A(t)^2 ∝ 1/ω(t) \,, 	
 \end{align}
 in addition to the irreversible changes associated with energy dissipation. 

\subsection{Frequency shift in Earth's gravity.}
\label{sec:gravity}

Piezoresistive cantilevers can be calibrated by measuring their bending in Earth’s gravity. Similarly, the oscillating cantilever exhibits a characteristic resonance frequency shift in Earth’s gravity. The reason for such a frequency shift is that each segment of the cantilever performs a "pendulum" motion in the external gravitational field. Left to itself, the length of the rope---or, in this case, the radius $R_n$ of the circle containing the oscillating trajectory of the segment of the cantilever---would determine the frequency of swinging oscillations. Instead, because the segments are coupled to each other by much stronger elastic forces, the swinging in a gravitational field leads to a small frequency shift, proportional to the ratio of gravitational energy $m_{\Lever}gL$ and the bending stiffness, Eq.~(\ref{eq:gravitational-shift-cantilever}). 

To describe this in better detail, we note that as each short segment on the cantilever moves perpendicular to the plane of the cantilever ($ζ_{n}(z,t)$, along $x$ in  Figure~\ref{fig:fig3}), it also moves along the length of the cantilever ($Ξ_{n}^{∥}(z,t)$,  along $z$ in  Figure~\ref{fig:fig3}) by a much smaller amount,  
	\begin{align}\label{eq:lengths2}
Ξ_{n}^{∥}(z)  ≈   -\frac{ ζ_n(z)^2 }{ 2R_n(z) }   \,.
	\end{align}
Geometrically, the curve $\{ ζ_n(z),  ζ_n(z)^2 /2R_n(z) \} $ describes a small arc of a circle of radius $R_n(z)$ with the center on the $z$-axis. Because the strain $ϵ_{zz}$ vanishes on the neutral surface, Eq.~(\ref{eq:epszz}), an arbitrary deformation of a thin cantilever leaves the length of a neutral surface unchanged. This property of neutral surface determines the "effective rope radius" $R_n(z)$ in Eq.~(\ref{eq:lengths2}) for arbitrary $ζ(z)$. Specifically, we label each point by its distance $s$ to the attachment point along the neutral surface of the bent cantilever, $0<s<L$, and can describe the bent shape of the cantilever parametrically with $\{x(s),z(s)\}$, subject to constraint $(dx/ds)^2+(dz/ds)^2=1$. The displacement $Ξ_{n}^{∥}(z)$ along the $z$ axis of segment of the cantilever can be represented as 
	\begin{align}\label{eq:lengths}
Ξ^{∥}(s) 
=& z(s)-s = ∫\limits_0^s ds' \para{  \frac{dz(s')}{ds'} -1 } \notag\\
=&  ∫\limits_0^s ds' \,\; \para{ \sqrt{ 1- {\der{ζ(s')}{s'}}^2 } -1}  \,.
	\end{align}
Expanding $Ξ^{∥}$ for small $ζ(s)$ gives equation Eq.~(\ref{eq:lengths2}) with effective rope radius 
	\begin{align}\label{eq:radius}
 \frac1{R_n(z)} = \frac1{ζ_n(z)^2} 
 ∫\limits_0^z dz' \bigpara{ \grad ζ_n(z')  }^2 \,,
	\end{align}
where we have relaxed the distinction between $z$ and $s$. For the fundamental mode, the rope radius $R_0(z)/z$ increases approximately linearly from $0.75$ at the base of the cantilever, $z=0$, to $0.86$ at the free end, $z=L$, 
\begin{align} \label{eq:arc-radius-0}
R_0(L) = 0.86L \,.	
\end{align}

Together, Eqs.~(\ref{eq:lengths2}), (\ref{eq:radius}), (\ref{eq:microscopic}), and (\ref{eq:normalization}) define the gravitational energy of the cantilever, bilinear-in-$Δθ$,  
	\begin{align}\label{eq:lever-gravity}
δ E_{\text{cantilever}}^{\text{gravity}} = &  μ g\cosθ × ∫\limits_0^L dz \; Ξ^{∥}(z) \notag\\
= & m_{\Lever} g\cosθ × b_nL ×  \frac{ Δθ^2}2 \;, 
	\end{align} 
where $m_{\Lever} = ρ A L$ is the total mass of the cantilever and $g\cosθ$ is the component of the Earth's gravity along the length of the cantilever. The numeric mode-dependent factor $b_n$ is  
\begin{align} 
	b_n = ∫_0^L  \frac{dz}L \para{ 1 - \frac{z}{L} }   \para{\grad ζ_n(z)}^2 	\,.
\end{align}
The gravitational shift of the fundamental mode,  $b_0≈0.21$, is (see Eqs.~(\ref{eq:sample-assembly}) and (\ref{eq:freqshift}) for comparison), 
\begin{align}\label{eq:gravitational-shift-cantilever}
	\para{\frac{Δω_0}{ω_0}}_{\Lever} =  0.21  \frac{  m_{\Lever} gL }{2\St_0} \, \cosθ \,.
\end{align} 

The sample attached at the end of the cantilever itself moves on a small arc of a circle of radius $R(L)$, Eq.~(\ref{eq:radius}). The  bilinear-in-$Δθ$ gravitational energy is  
\begin{align} \label{eq:sample-gravity}
δ E_{\text{Sample}}^{\text{gravity}} =  m_{\Sample}  g\cosθ × R_n(L) ×  \frac{Δθ^2}{2}	\,.
\end{align}
For the fundamental mode, Eq.~(\ref{eq:arc-radius-0}), 
\begin{align}\label{eq:gravitational-shift-sample}
\para{\frac{Δω_0}{ω_0}}_{\Sample} =  0.86\;  \frac{m_{\Sample} g L  }{ 2\St_0} \, \cosθ   \,. 	
\end{align}

\subsection{ Effects of non-uniform magnetic field} 
\label{sec:non-uniform-magnetic-field}

A non-uniform magnetic field can shift the resonance frequency of the cantilever through several mechanisms. Unless the experiment is specifically designed to observe these effects, the resonance shifts resulting from a non-uniform magnetic field are much smaller than the frequency shifts associated with sample rotation in the magnetic field. Typically, they are smaller by a factor proportional to the first or second power of the ratio of the cantilever's length to the size of the coil generating the applied magnetic field (which determines the size of the gradients of magnetic field). Here we provide a brief semi-quantitative discussion of the frequency shifts associated with spatial gradients and spatial curvatures of magnetic field.

Small changes in the free energy of the sample in a non-uniform magnetic field 
\begin{align} \label{eq:nonuniform}
dF = - \M ⋅ d\B 	
\end{align}
originate from the finite extent $ΔX$ of the trajectory of the sample at the free end of the cantilever (Eq.~(\ref{eq:deltaX}) for the fundamental mode) along the $x$-axis (Figure~\ref{fig:fig3}) as well as the pendulum-like motion of the sample along the $z$-axis, Eq.~(\ref{eq:lengths2}). The frequency shifts associated with these two orthogonal motions have different characters. 

The field gradient along the length of the cantilever acts on the sample in the same way as a force of gravity, Eq.~(\ref{eq:sample-gravity}). The sample experiences a uniform force along the length of the cantilever~:
\begin{align}
F_z = -\frac{dF}{dz} = \M ⋅ \frac{d\B}{dz} 	
\end{align}
The associated frequency shift is obtained from Eq.~(\ref{eq:gravitational-shift-sample}) by replacing $m_{\Sample} g\cosθ$ with $\M ⋅ \fra{d\B}{dz} 	$. 
For example, the gravitational frequency shift of a sample with 1 $μ_{\Bohr}$/f.u magnetization and an atomic number of 100/f.u. is about 100 times smaller than the frequency shift in magnetic field gradient of 0.1 T/cm. 

The inhomogeneity of magnetic field along the $x$-axis introduces a parabolic potential that depends on both gradients and spatial curvature of the inhomogeneous  magnetic field, 
\begin{align}\label{eq:spatial-field-curvature}
	ΔF  =  \frac{ΔX^2}{2}\bigpara{ 
	 - \M_{μ} \frac{d^2\B_{μ}}{dx^2}   - \hat{χ}_{μν}  \frac{d\B_{ν}}{dx} \frac{d\B_{μ}}{dx} 
	 } 	\,,
\end{align}
where $\hat{χ}_{μν}   = \fra{d\M_{μ}}{d\B_{ν}}$ is the magnetic susceptibility.  Equation~(\ref{eq:spatial-field-curvature}) produces a small $Δ X^2$ term in Eq.~(\ref{eq:EofX}), and through Eq.~(\ref{eq:deltaX}), a small $Δ θ^2$ term in Eq.~(\ref{eq:effective0}). For the fundamental mode, the associated frequency shift is given by Eq.~(\ref{eq:freqshift}) with parenthesis factor in  Eq.~(\ref{eq:spatial-field-curvature})  replacing the magnetotropic coefficient. 

For a cantilever, where oscillations rotate as well as displace the sample simultaneously, Eq.~(\ref{eq:effectivelength}), the frequency shift associated with the spatially non-uniform magnetic field is suppressed compared to ``rotational shift'', Eq.~(\ref{eq:freqshift}), by a factor of the square of the ratio of the size of the cantilever to the size of the magnet. We note, however, that the two shifts have different angular dependence as well as different dependence on components of the magnetization and magnetic susceptibility.


\section{Mechanics of an oscillating cantilever}
\label{sec:cantilever}

\subsection{ Thin cantilever and arc stiffness.}

\begin{figure}[t!!!]
				\centerline{  
				\includegraphics[width=0.8\columnwidth]{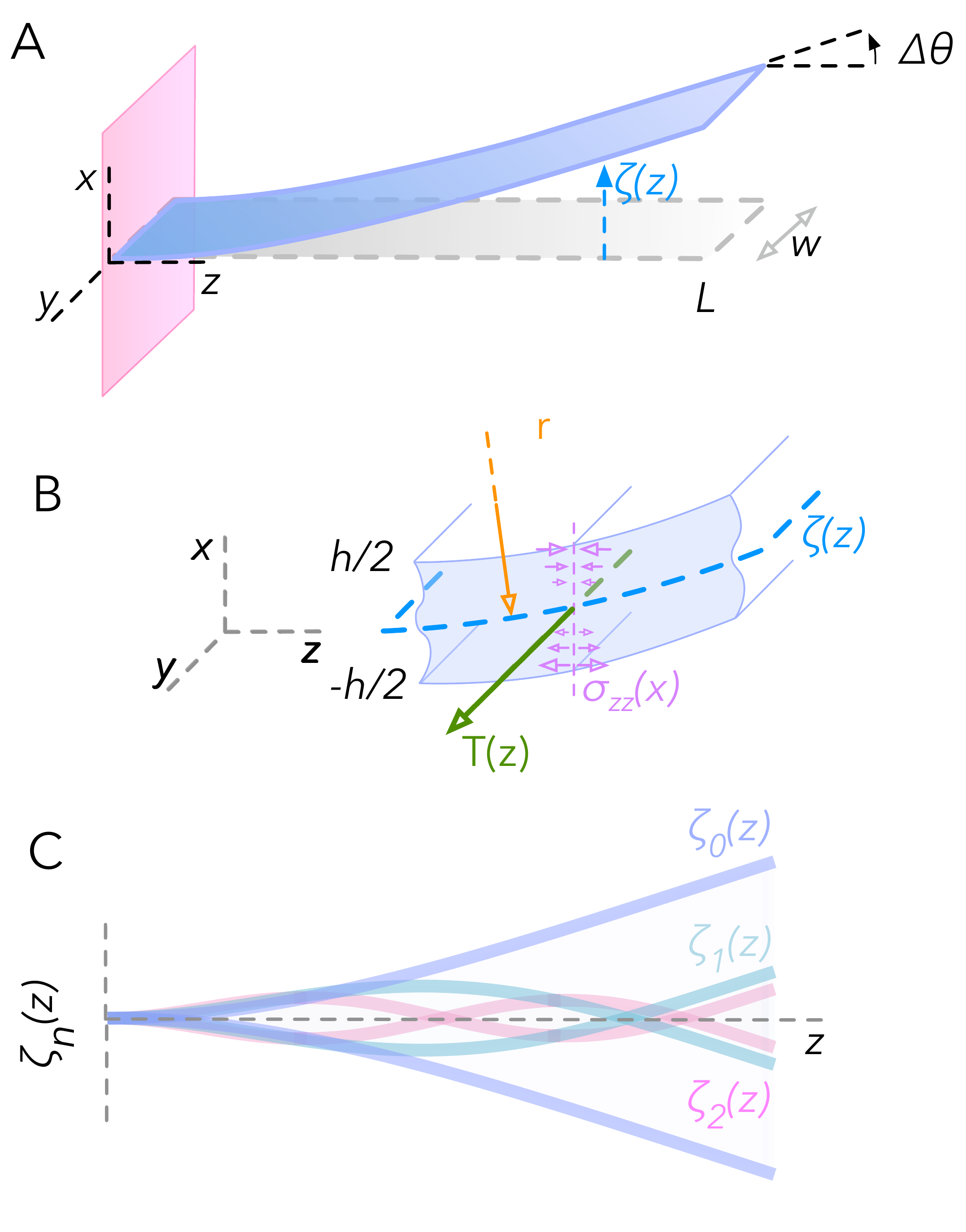}  }
				\caption{ %
Resonating cantilever.  	 
A. Schematic of an oscillating cantilever. The angle of $∆θ$ at the tip of the cantilever (z=L) is equal to the gradient of the displacement, $∆θ = \grad ζ(z) $. 
B. Cross-section of the cantilever. The blue dashed line indicates a neutral surface. The purple arrow indicates the magnitude of $σ_{zz}$ that changes sign across the neutral surface. The orange line indicates the radius of curvature $1/r(z) = \lap ζ(z)$. The green arrow indicates the cross-sectional torque in the cantilever, Eq.~(\ref{eq:crossectional-torque}). 
C. Plot of the shape of the cantilever in fundamental (blue) and the next two modes of the cantilever, Eqs.~(\ref{eq:microscopic}) and (\ref{eq:normalization}). 
}
				\label{fig:fig3} 
				\end{figure}

For completeness, we discuss here the mechanics of a thin cantilever  \cite{Landau-Elasticity1986}. Throughout this discussion, we shall assume that the thickness of the cantilever is much smaller than either of its lateral dimensions, $h ≪ L,w$ (Figure \ref{fig:fig3}). In such  a thin-plate approximation, the bent state of the cantilever is described completely by the $x$-axis displacement of the cantilever at a distance $z$ from the point of attachment, $ζ(z)$. The mechanical energy of the cantilever is a bilinear function of the displacement $ζ(z)$, 
	\begin{align}\label{eq:totalenergy}
E_{\text{tot}} 
= \frac{μ}{2} ∫\limits_0^L\!\!dz{\der{ζ(z,t)}{t}}^2  
+ \frac{γ}{2}∫\limits_0^L\!\!dz \para{\frac{ d^2ζ(z,t) }{dz^2} }^2 \!.
	\end{align}
The mechanical energy of the cantilever determines the resonant frequencies and the effective bending stiffnesses. The first term in Eq.~(\ref{eq:totalenergy}) is the kinetic energy of the oscillating cantilever, proportional to the square of the velocity of each segment of the cantilever,  $μ v(z)/2=μ [d{ζ(z)}/{dt}]^2/2$. Here $μ$ is the mass of the lever per unit length, $μ = ρ A$ where $A=w h$ is the cross-sectional area of the cantilever, $w$ is the width of the cantilever, and $h$ is its thickness. $ρ$ is the density of the cantilever. 

The second term in Eq.~(\ref{eq:totalenergy}) is the elastic energy,  proportional to the second derivative---curvature---of $ζ(z)$ for each segment of the cantilever,  $1/r(z) = \fra{∂^2ζ(z)}{∂ z^2}$, where $r(z)$ is the local radius of curvature at $z$. 

The arc stiffness $γ$ in Eq.~(\ref{eq:totalenergy}) determines the elastic energy of the cantilever for arbitrary $ζ(z)$. The arc stiffness $γ$ is itself determined by the elastic deformation inside the cantilever. For small deformations of a thin cantilever, only one component, $σ_{zz}$, of the stress tensor is nonzero, and it alone determines $γ$. This is because the perpendicular component of the elastic stress on the free surface of the cantilever is zero which requires that all stress components except $σ_{zz}$ must vanish there. For a thin cantilever, this implies that all stress components, with the exception of $σ_{zz}$, are zero, not only on the surface of but also inside of the cantilever.\cite{Landau-Elasticity1986} 

The stress  $σ_{zz}$ is a linear function of the distance to the "neutral surface" and vanishes on it, $σ_{zz} = b(z) x$, where $x$ runs from $-h/2$ to $h/2$ (Figure~(\ref{fig:fig3}). For a thin cantilever, the neutral surface is in the middle of the cross-sectional area.  $b(z)$ can be found from elastic equations that connect the strains  (described via $ζ(z)$) to the stresses. Assuming that the cantilever is cut out of a cubic crystal parallel to its crystallographic directions, the elastic equations are  
	\begin{align}\label{eq:el-isotropic}
σ_{xx}+σ_{yy}+σ_{zz} =&\; (c_{11}+2c_{12}) ×  (ϵ_{xx}+ϵ_{yy}+ϵ_{zz}) \,,   \notag\\
σ_{xx}-σ_{yy} =& \;(c_{11} - c_{12}) × \;\; (ϵ_{xx}-ϵ_{yy}) 				                \,, \notag\\
σ_{xx}+σ_{yy}-2σ_{zz} =& \;(c_{11} - c_{12}) × \;\; (ϵ_{xx}+ϵ_{yy}-2ϵ_{zz}) \,,  \notag\\
σ_{i≠ j}=& \;c_{66}\;\; ϵ_{i≠ j} \,.
	\end{align}
	Here, only one, $σ_{zz}$, component of a stress tensor is nonzero. Therefore, only compressional strains are non-zero, 
	\begin{align}\label{eq:elastic1}
ϵ_{zz}=&\frac1Y σ_{zz}\,, \qquad 
ϵ_{xx}=ϵ_{yy}=-σ ϵ_{zz} = -\frac{σ}{Y} σ_{zz} \,.
	\end{align}
where  $σ$ is the  Poisson's ratio and $Y$ is the Young's modulus of the cubic crystal, 
	\begin{align}
Y=\frac{\para{c_{11}-c_{12}}  \para{c_{11}+2 c_{12}} }{ c_{11}+c_{12} }\;, \qquad
σ = \frac{c_{12}}{ c_{11}+ c_{12}} \,.
	\end{align}

Equations~(\ref{eq:elastic1}) show that  the strains $ϵ_{xx}$, $ϵ_{yy}$, and $ϵ_{zz}$ are all linear in $x$ because $σ_{zz}(x)=b(z)x$ is linear in $x$. The linear-in-$x$  strain $ϵ_{zz} = x/r(z)$ implies that the shape of the bent cantilever is an arc of a thin annulus. This is because under such deformation each vertical line $z=\text{const}$ will deform into radial segments and each horizontal line $x=\text{const}$ will deform into an arc of a circle, with the same center as the arc of the neutral surface (Figure~\ref{fig:fig3}). 
$r(z)$ is the radius of curvature of a neutral surface  (Figure \ref{fig:fig3})~:
	\begin{align} \label{eq:radiusofcurv}
\frac1{r(z)} = \lap ζ \,,
	\end{align}
where $\grad= \fra{∂}{∂ z}$ denotes the derivative with respect to $z$. This establishes the connection between the strain and the bent shape of the cantilever, 
\begin{align}\label{eq:epszz}
ϵ_{zz}(x,z) = x\; \lap ζ(z) \,. 	
\end{align}
Elastic equations, Eq.~(\ref{eq:elastic1}), in turn, require that the coefficient $b(z)$ in the stress $σ_{zz}(x,z) = b(z) x$ is equal to $b(z) = Y/r(z) = Y \lap ζ(z) $, and, therefore, 
\begin{align}\label{eq:sigmazz}
	σ_{zz}(x,z) = x\; Y\, \lap ζ(z) \,.
\end{align}

The arc stiffness $γ$ in Eq.~(\ref{eq:totalenergy})  determines the elastic energy of the cantilever per unit length.  We can integrate the elastic energy of the cantilever in a small volume, $dE = (\sfrac12) ϵ_{ij}σ_{ij} dV$, 
	\begin{align}
	\frac{Δ E}{Δ V} 
	= \frac12 ϵ_{zz}σ_{zz}  
	=  \frac1{2Y} σ_{zz}^2   
	= \frac{Y}{2} \; \bigpara{ \lap ζ}^2 x^2  \,,
	\end{align}
over the cross-sectional area to obtain the elastic energy of the cantilever per unit length,
	\begin{align}\label{eq:energyperunitlength}
\frac{Δ E}{Δ z} 
	= \frac{ γ }{2} \; \bigpara{ \lap ζ}^2 
	  \,.
	\end{align}
which defines the arc stiffness, 
\begin{align}\label{eq:arcstiffness}
γ = Y wh \mi = Ywh^3/12 \,,
\end{align}
in terms of geometric and mechanical parameters of the cantilever (Figure \ref{fig:fig3}). Here $\mi = (1/h)∫ x^2 dx$ is the average $x^2$ over the cross-section of the cantilever. $\mi w h$  is the geometric ``moment of inertia" of the cross-sectional area of the cantilever \cite{Landau-Elasticity1986}. For a thin cantilever, $\mi = h^2/12$. 

The boundary conditions for the equations of motion follow from an expression for the mechanical energy of a non-uniform cantilever for which $w(z)$,  $Y(z)$, $ρ(z)$, and $h(z)$ vary along its length, 
	\begin{align}\label{eq:totalenergy-arb}
E_{\text{tot}} 
	= & \frac{1}{2}\!∫\limits_0^L\!\!dz\, μ(z) \! \para{\!\frac{dζ(z,t)}{dt}\!}^2  
\!\!\!+\! \frac{1}{2}\!∫\limits_0^L\!\!dz\,  γ(z) \!\para{\! \frac{d^2 ζ(z,t) }{ dz^2 }\!}^2 \!,
	\end{align}
	where $μ(z)= w(z) h(z) ρ(z)$, and $γ(z) = Y(z) w(z) h(z)^3/12$.  
For example, if the cantilever consists of two unequal uniform segments such that $w(z)$ and/or $h(z)$ have a sharp step, the equations of motion require the continuity of $ζ(z)$,  $\grad ζ(z)$,  $γ(z) \grad^2 ζ(z) $, and $γ(z) \grad^3 ζ(z)$ across the step. 

When the cantilever is bent locally into an arc of curvature $\lap ζ(z)$,  a finite cross-sectional torque $T(z)$ is developed in its cross-section (Figure \ref{fig:fig3}). The cross-sectional torque $T(z)$ is equal to the change in elastic energy of the cantilever, Eq.~(\ref{eq:totalenergy}), when local curvature of the cantilever is varied,  $δ(\fra{dE}{dz}) =  T(z) δ(\lap ζ)$. Equation~(\ref{eq:totalenergy}) gives $T(z) =  γ \lap ζ(z)$. Therefore, the arc stiffness $γ$ in Eq.~(\ref{eq:totalenergy}) has the meaning of the coefficient of proportionality between crossectional torque $T(z)$ and the local curvature of the cantilever.  
The same result is found by direct integration of the stress $σ_{zz}(x)$ in Eq.~(\ref{eq:sigmazz}) over the cross-sectional area of the cantilever, 
 	\begin{align}\label{eq:crossectional-torque}
T(z) = & ∫ x σ_{zz}(x) dA = γ \lap ζ(z)  \,, 
	\end{align} 
where arc stiffness $γ$ is given by Eq.~(\ref{eq:arcstiffness}).

\subsection{Oscillating cantilever.}

The motion of the cantilever near one of its mechanical resonances is determined by the equations of motion for $ζ(z,t)$,
	\begin{align}\label{eq:Bilaplacian}
\frac{d^2ζ(z,t)}{dt^2}  + \frac{γ}{μ} \grad^4ζ(z,t)  =0 \,,
	\end{align}
which follow from the energy functional in Eq.~(\ref{eq:totalenergy}) \cite{Landau-Mechanics1976}. Near the resonance, the cantilever oscillates at a single frequency $ω$, 
	\begin{align}
ζ(z,t)= ζ_n(z) e^{iω_n t}\,,
	\end{align}
determined by one of the eigenvalues of the Bilaplacian 
	\begin{align} \label{eq:omega-kappa}
\grad^4 ζ_n(z) = \frac{κ_n^4}{L^4} \; ζ_n(z) \;, \qquad
ω_n^2 =\frac{κ_n^4}{L^4} \; \frac{γ}{μ}     \,,
	\end{align}
subject to the boundary conditions of the cantilever. For a cantilever clamped at one end, $z=0$, and free at the other, $z=L$, 	\begin{align}\label{eq:boundary}
	\grad^2ζ_n(L) = 0 \,, \;
	\grad^3ζ_n(L) = 0 \,, \;
	\grad ζ_n(0) = 0, \;
	ζ_n(0) = 0 \,. 
	\end{align}
The eigenfunctions of the Bilaplacian in Eq.~(\ref{eq:omega-kappa}) is solved with  
	\begin{align}\label{eq:superpos}
ζ_n(z)\!=\!A\cos\frac{κ_n z}{L} \!+\! B\sin\frac{κ_n z}{L}\!+\!C\cosh\frac{κ_n z}{L}\!+\! D\sinh\frac{κ_n z}{L} \,,
	\end{align}
where $A,B,C,D$ are constant factors determined by the boundary conditions, 
	\begin{align}\label{eq:clampedcantilever}
-A\cosκ_n - B\sinκ_n + C\coshκ_n + D\sinhκ_n =& \; 0 \notag\\
A\sinκ_n - B\cosκ_n + C\sinhκ_n + D\coshκ_n =& \; 0 \notag\\
B+D  =&  \; 0 \notag\\
A+C=& \; 0 \,.
	\end{align}
The eigenvalues $κ_n$ of the Bilaplacian are determined by the condition that the matrix in Eq.~(\ref{eq:clampedcantilever}) has zero determinant, 
	\begin{align}
	\text{Det} \matr{
 -\cosκ_n & -\sinκ_n & \coshκ_n & \sinhκ_n \\
 \sinκ_n & -\cosκ_n & \sinhκ_n & \coshκ_n \\
 0 & 1 & 0 & 1 \\
 1 & 0 & 1 & 0 
} =0 \,,
	\end{align}
which simplifies to
	\begin{align}\label{eq:det-clampedcantilever}
	1+\cosκ_n\coshκ_n=0 \,.
	\end{align}
Finally, we obtain 
	\begin{align}\label{eq:sequence}
	κ_0 ≈  0.597π, \qquad  	κ_{n=1,2,...} ≈ (n+\sfrac12)π \,.
	\end{align}
The set of resonant frequencies of the cantilever is now determined via Eq.~(\ref{eq:omega-kappa}), 
\begin{align}\label{eq:omega-n}
ω_n = \frac{κ_n^2}{L^2} \; \sqrt{\frac{γ}{μ}} \,.	
\end{align}
The frequency of the lowest frequency (fundamental) mode of a thin cantilever with a uniform  cross-section is given by combining Eqs.~(\ref{eq:omega-kappa}, \ref{eq:sequence}, \ref{eq:arcstiffness}),
	\begin{align}\label{eq:lowestmode}
ω_0^{} =1.02 \frac{h}{L^2} \sqrt{\frac{Y}{ρ}}\,, 
	\end{align}
	where the numerical factor, $(0.597π)^2/\sqrt{12} ≈ 1.02$.

The bending shape of the cantilever near the resonance  $ζ_n(z)$ is found from Eq.~(\ref{eq:clampedcantilever}), 
	\begin{align} 
	A=& -C = \coshκ_n + \cosκ_n , \notag\\ 
	B =& -D = \sinκ_n - \sinhκ_n \,,
	\end{align}
	and Eq.~(\ref{eq:superpos}), 
	\begin{align}\label{eq:microscopic}
	&ζ_n(z) = L\; N_n\Big[ 
	\para{\coshκ_n + \cosκ_n }\para{\cosh\frac{κ_nz}{L} -\cos\frac{κ_nz}{L} } \notag\\
	&\qquad  -\para{\sinhκ_n - \sinκ_n}\para{ \sinh\frac{κ_nz}{L} - \sin\frac{κ_nz}{L} }
	\Big] \,.
	\end{align}
The normalization factor $N_n$ can be chosen to fix the bending angle $∆θ(t) = \grad ζ(z=L,t)$ at the free end of the cantilever, $z=L$. We parametrize the motion of the cantilever near the resonance as $ζ_n(z,t) = ζ_n(z) ∆θ(t)$ where  $ζ_n(z)$ is normalized to have a unit slope at the free end, $\grad ζ_n(z=L) =1$,  and $∆θ(t)$ oscillates at the resonance frequency, $∆θ(t)= ∆θ e^{iω_nt} $,  
	\begin{align}\label{eq:normalization}
	N_n = \frac{1}{2κ_n}  ×  \frac{1}{ 	\coshκ_n\sinκ_n + \cosκ_n \sinhκ_n}  \,.
	\end{align} 
The displacement $∆X_n=ζ_n(L)$ and the angle $∆θ_n = \grad ζ_n(L)$ at the free end of the cantilever are proportional to each other, 
\begin{align}\label{eq:effectivelength}
	∆X_n  = \frac{ 1}{ κ_n  (\cotκ_n + \coth κ_n) }\, L\,∆θ_n   \,. 	
\end{align} 
This gives $∆X_0  = 0.73L \, ∆θ_0$ for the fundamental mode, $n=0$, and $0.21, 0.13, ⋯ $ for $n=1,2, ⋯$.

\subsection{The effective bending stiffness $\St_n$ in the vicinity of a resonance.}

Together, Eqs.~(\ref{eq:microscopic}), (\ref{eq:normalization}), and (\ref{eq:totalenergy}) describe the  kinetic and potential energy of the cantilever near the $n$-th resonance mode in terms of the rotation angle at the tip of the cantilever, $∆θ(t)= ∆θ e^{iω_nt} $ ~: 
	\begin{align}\label{eq:effective}
  E= \frac{I_n}2 {\der{∆θ}{t}}^2 + \frac{\St_n}2 ∆θ^2. 
	\end{align} 
The values of $\St_n$ and $I_n$ are evaluated by substituting the shape of the cantilever in the  $n$-th resonance mode,  Eqs.~(\ref{eq:microscopic}), and (\ref{eq:normalization}), into the kinetic and elastic energy terms in Eq.~(\ref{eq:totalenergy}) and expressing both as bilinear functions of the angle $∆θ$. Calculating the kinetic energy  in Eq.~(\ref{eq:totalenergy}),
	\begin{align}\label{eq:effectiveIintegral}
{\der{∆θ}{t}}^2 ∫\limits_0^L dz  {ζ_n(z)}^2 = L^3 \para{\frac{a_n}{κ_n^3}} \, {\der{∆θ}{t}}^2 \,,
	\end{align}
where $a_{n=0,1,2,...} =\curly{ 0.870, 1.13, 1.97, ... }$ is a mode-dependent numeric factor defined by the average square of the displacement of the cantilever,  
	\begin{align}\label{eq:def-an}
a_n = {\frac{κ_n^3}{L^3}} ∫\limits_0^L dz  \bigpara{ζ_n(z)}^2 \,,
	\end{align}
with $ζ_n(z)$ from Eq.~(\ref{eq:microscopic}) and normalization factor $N_n$ from Eq.~(\ref{eq:normalization}) where $∆θ$ is set to unity. We obtain the parameter $I_n$ in Eq.~(\ref{eq:effective}), 
	\begin{align}\label{eq:effectiveI}
 I_n = \frac{a_n}{κ_n^3}  × μ L^3 
 		= \frac{a_n}{κ_n^3} ×  ρ whL^3 \,.
	\end{align}
The integral in the elastic energy in Eq.~(\ref{eq:totalenergy}) can be reduced to the one in Eq.~(\ref{eq:def-an}) using equations of motion,  Eq.~(\ref{eq:Bilaplacian}),  and we obtain
	\begin{align}\label{eq:effectiveK}
  \St_n =   a_n κ_n ×  \frac{γ }{ L }  =  a_n κ_n ×  \frac{Ywh^3}{12 L }   \,. 
	\end{align} 
Equations~(\ref{eq:effectiveI}) and (\ref{eq:effectiveK}), through their definition in  Eq.~(\ref{eq:effective}), produce the correct resonance frequency, Eq.~(\ref{eq:omega-n}). 

\phantom{ }

\begin{acknowledgments} 
	{\it Acknowledgments.} The work at the Los Alamos National Laboratory is supported by National Science Foundation Cooperative Agreement DMR-1157490, DMR-1644779, the state of Florida, and the U.S. Department of Energy. A.S. acknowledges support from the DOE/BES “Science of 100T” grant. B.J.R. acknowledges funding from the National Science Foundation under grant no. DMR-1752784. 
\end{acknowledgments}

\hypersetup{linkcolor=black,citecolor=black,filecolor=black,urlcolor=black}

\bibliographystyle{apsrev4-1}

\begin{thebibliography}{25}%
\makeatletter
\providecommand \@ifxundefined [1]{%
 \@ifx{#1\undefined}
}%
\providecommand \@ifnum [1]{%
 \ifnum #1\expandafter \@firstoftwo
 \else \expandafter \@secondoftwo
 \fi
}%
\providecommand \@ifx [1]{%
 \ifx #1\expandafter \@firstoftwo
 \else \expandafter \@secondoftwo
 \fi
}%
\providecommand \natexlab [1]{#1}%
\providecommand \enquote  [1]{``#1''}%
\providecommand \bibnamefont  [1]{#1}%
\providecommand \bibfnamefont [1]{#1}%
\providecommand \citenamefont [1]{#1}%
\providecommand \href@noop [0]{\@secondoftwo}%
\providecommand \href [0]{\begingroup \@sanitize@url \@href}%
\providecommand \@href[1]{\@@startlink{#1}\@@href}%
\providecommand \@@href[1]{\endgroup#1\@@endlink}%
\providecommand \@sanitize@url [0]{\catcode `\\12\catcode `\$12\catcode `\&12\catcode `\#12\catcode `\^12\catcode `\_12\catcode `\%12\relax}%
\providecommand \@@startlink[1]{}%
\providecommand \@@endlink[0]{}%
\providecommand \url  [0]{\begingroup\@sanitize@url \@url }%
\providecommand \@url [1]{\endgroup\@href {#1}{\urlprefix }}%
\providecommand \urlprefix  [0]{URL }%
\providecommand \Eprint [0]{\href }%
\providecommand \doibase [0]{http://dx.doi.org/}%
\providecommand \selectlanguage [0]{\@gobble}%
\providecommand \bibinfo  [0]{\@secondoftwo}%
\providecommand \bibfield  [0]{\@secondoftwo}%
\providecommand \translation [1]{[#1]}%
\providecommand \BibitemOpen [0]{}%
\providecommand \bibitemStop [0]{}%
\providecommand \bibitemNoStop [0]{.\EOS\space}%
\providecommand \EOS [0]{\spacefactor3000\relax}%
\providecommand \BibitemShut  [1]{\csname bibitem#1\endcsname}%
\let\auto@bib@innerbib\@empty
\bibitem [{\citenamefont {Bishop}\ and\ \citenamefont {Reppy}(1978)}]{Bishop1978}%
  \BibitemOpen
  \bibfield  {author} {\bibinfo {author} {\bibfnamefont {D.~J.}\ \bibnamefont {Bishop}}\ and\ \bibinfo {author} {\bibfnamefont {J.~D.}\ \bibnamefont {Reppy}},\ }\href@noop {} {\bibfield  {journal} {\bibinfo  {journal} {Phys. Rev. Lett.}\ }\textbf {\bibinfo {volume} {40}},\ \bibinfo {pages} {1727} (\bibinfo {year} {1978})}\BibitemShut {NoStop}%
\bibitem [{\citenamefont {Kleiman}\ \emph {et~al.}(1984)\citenamefont {Kleiman}, \citenamefont {Bishop}, \citenamefont {Pindak},\ and\ \citenamefont {Taborek}}]{Kleiman1984}%
  \BibitemOpen
  \bibfield  {author} {\bibinfo {author} {\bibfnamefont {R.~N.}\ \bibnamefont {Kleiman}}, \bibinfo {author} {\bibfnamefont {D.~J.}\ \bibnamefont {Bishop}}, \bibinfo {author} {\bibfnamefont {R.}~\bibnamefont {Pindak}}, \ and\ \bibinfo {author} {\bibfnamefont {P.}~\bibnamefont {Taborek}},\ }\href@noop {} {\bibfield  {journal} {\bibinfo  {journal} {Phys. Rev. Lett.}\ }\textbf {\bibinfo {volume} {53}},\ \bibinfo {pages} {2137} (\bibinfo {year} {1984})}\BibitemShut {NoStop}%
\bibitem [{\citenamefont {Kleiman}\ \emph {et~al.}(1985)\citenamefont {Kleiman}, \citenamefont {Kaminsky}, \citenamefont {Reppy}, \citenamefont {Pindak},\ and\ \citenamefont {Bishop}}]{Kleiman1985}%
  \BibitemOpen
  \bibfield  {author} {\bibinfo {author} {\bibfnamefont {R.~N.}\ \bibnamefont {Kleiman}}, \bibinfo {author} {\bibfnamefont {G.~K.}\ \bibnamefont {Kaminsky}}, \bibinfo {author} {\bibfnamefont {J.~D.}\ \bibnamefont {Reppy}}, \bibinfo {author} {\bibfnamefont {R.}~\bibnamefont {Pindak}}, \ and\ \bibinfo {author} {\bibfnamefont {D.~J.}\ \bibnamefont {Bishop}},\ }\href {\doibase 10.1063/1.1138425} {\bibfield  {journal} {\bibinfo  {journal} {Rev. Sci. Instr.}\ }\textbf {\bibinfo {volume} {56}},\ \bibinfo {pages} {2088} (\bibinfo {year} {1985})}\BibitemShut {NoStop}%
\bibitem [{\citenamefont {Kleiman}\ \emph {et~al.}(1987)\citenamefont {Kleiman}, \citenamefont {Agnolet},\ and\ \citenamefont {Bishop}}]{Kleiman1987}%
  \BibitemOpen
  \bibfield  {author} {\bibinfo {author} {\bibfnamefont {R.~N.}\ \bibnamefont {Kleiman}}, \bibinfo {author} {\bibfnamefont {G.}~\bibnamefont {Agnolet}}, \ and\ \bibinfo {author} {\bibfnamefont {D.~J.}\ \bibnamefont {Bishop}},\ }\href@noop {} {\bibfield  {journal} {\bibinfo  {journal} {Phys. Rev. Lett.}\ }\textbf {\bibinfo {volume} {59}},\ \bibinfo {pages} {2079} (\bibinfo {year} {1987})}\BibitemShut {NoStop}%
\bibitem [{\citenamefont {Worthington}\ \emph {et~al.}(1987)\citenamefont {Worthington}, \citenamefont {Gallagher},\ and\ \citenamefont {Dinger}}]{Worthington1987}%
  \BibitemOpen
  \bibfield  {author} {\bibinfo {author} {\bibfnamefont {T.~K.}\ \bibnamefont {Worthington}}, \bibinfo {author} {\bibfnamefont {W.~J.}\ \bibnamefont {Gallagher}}, \ and\ \bibinfo {author} {\bibfnamefont {T.~R.}\ \bibnamefont {Dinger}},\ }\href@noop {} {\bibfield  {journal} {\bibinfo  {journal} {Phys. Rev. Lett.}\ }\textbf {\bibinfo {volume} {59}},\ \bibinfo {pages} {1160} (\bibinfo {year} {1987})}\BibitemShut {NoStop}%
\bibitem [{\citenamefont {Bishop}\ \emph {et~al.}(1992)\citenamefont {Bishop}, \citenamefont {Gammel}, \citenamefont {Huse},\ and\ \citenamefont {Murray}}]{Bishop1992}%
  \BibitemOpen
  \bibfield  {author} {\bibinfo {author} {\bibfnamefont {D.~J.}\ \bibnamefont {Bishop}}, \bibinfo {author} {\bibfnamefont {P.~L.}\ \bibnamefont {Gammel}}, \bibinfo {author} {\bibfnamefont {D.~A.}\ \bibnamefont {Huse}}, \ and\ \bibinfo {author} {\bibfnamefont {C.~A.}\ \bibnamefont {Murray}},\ }\href@noop {} {\bibfield  {journal} {\bibinfo  {journal} {Science}\ }\textbf {\bibinfo {volume} {255}},\ \bibinfo {pages} {165} (\bibinfo {year} {1992})}\BibitemShut {NoStop}%
\bibitem [{\citenamefont {Bolle}\ \emph {et~al.}(1999)\citenamefont {Bolle}, \citenamefont {Aksyuk}, \citenamefont {F.Pardo}, \citenamefont {Gammel}, \citenamefont {Zeldov}, \citenamefont {Bucher}, \citenamefont {Boie}, \citenamefont {Bishop},\ and\ \citenamefont {Nelson}}]{Bolle1999}%
  \BibitemOpen
  \bibfield  {author} {\bibinfo {author} {\bibfnamefont {C.~A.}\ \bibnamefont {Bolle}}, \bibinfo {author} {\bibfnamefont {V.}~\bibnamefont {Aksyuk}}, \bibinfo {author} {\bibnamefont {F.Pardo}}, \bibinfo {author} {\bibfnamefont {P.~L.}\ \bibnamefont {Gammel}}, \bibinfo {author} {\bibfnamefont {E.}~\bibnamefont {Zeldov}}, \bibinfo {author} {\bibfnamefont {E.}~\bibnamefont {Bucher}}, \bibinfo {author} {\bibfnamefont {R.}~\bibnamefont {Boie}}, \bibinfo {author} {\bibfnamefont {D.~J.}\ \bibnamefont {Bishop}}, \ and\ \bibinfo {author} {\bibfnamefont {D.~R.}\ \bibnamefont {Nelson}},\ }\href@noop {} {\bibfield  {journal} {\bibinfo  {journal} {Nature}\ }\textbf {\bibinfo {volume} {399}},\ \bibinfo {pages} {43} (\bibinfo {year} {1999})}\BibitemShut {NoStop}%
\bibitem [{\citenamefont {Chiaverini}\ \emph {et~al.}(2001)\citenamefont {Chiaverini}, \citenamefont {Yasumura},\ and\ \citenamefont {Kapitulnik}}]{Chiaverini2001}%
  \BibitemOpen
  \bibfield  {author} {\bibinfo {author} {\bibfnamefont {J.}~\bibnamefont {Chiaverini}}, \bibinfo {author} {\bibfnamefont {K.}~\bibnamefont {Yasumura}}, \ and\ \bibinfo {author} {\bibfnamefont {A.}~\bibnamefont {Kapitulnik}},\ }\href@noop {} {\bibfield  {journal} {\bibinfo  {journal} {Physical Review B}\ }\textbf {\bibinfo {volume} {64}},\ \bibinfo {pages} {014516} (\bibinfo {year} {2001})}\BibitemShut {NoStop}%
\bibitem [{\citenamefont {Modic}\ \emph {et~al.}(2018)\citenamefont {Modic}, \citenamefont {Bachmann}, \citenamefont {Ramshaw}, \citenamefont {Arnold}, \citenamefont {Shirer}, \citenamefont {Estry}, \citenamefont {Betts}, \citenamefont {Ghimire}, \citenamefont {Bauer}, \citenamefont {Schmidt}, \citenamefont {Baenitz}, \citenamefont {Svanidze}, \citenamefont {McDonald}, \citenamefont {Shekhter},\ and\ \citenamefont {Moll}}]{Modic2018a}%
  \BibitemOpen
  \bibfield  {author} {\bibinfo {author} {\bibfnamefont {K.~A.}\ \bibnamefont {Modic}}, \bibinfo {author} {\bibfnamefont {M.~D.}\ \bibnamefont {Bachmann}}, \bibinfo {author} {\bibfnamefont {B.~J.}\ \bibnamefont {Ramshaw}}, \bibinfo {author} {\bibfnamefont {F.}~\bibnamefont {Arnold}}, \bibinfo {author} {\bibfnamefont {K.~R.}\ \bibnamefont {Shirer}}, \bibinfo {author} {\bibfnamefont {A.}~\bibnamefont {Estry}}, \bibinfo {author} {\bibfnamefont {J.~B.}\ \bibnamefont {Betts}}, \bibinfo {author} {\bibfnamefont {N.~J.}\ \bibnamefont {Ghimire}}, \bibinfo {author} {\bibfnamefont {E.~D.}\ \bibnamefont {Bauer}}, \bibinfo {author} {\bibfnamefont {M.}~\bibnamefont {Schmidt}}, \bibinfo {author} {\bibfnamefont {M.}~\bibnamefont {Baenitz}}, \bibinfo {author} {\bibfnamefont {E.}~\bibnamefont {Svanidze}}, \bibinfo {author} {\bibfnamefont {R.~D.}\ \bibnamefont {McDonald}}, \bibinfo {author} {\bibfnamefont {A.}~\bibnamefont {Shekhter}}, \ and\ \bibinfo {author} {\bibfnamefont {P.~J.~W.}\ \bibnamefont {Moll}},\ }\href {\doibase 10.1038/s41467-018-06412-w} {\bibfield  {journal} {\bibinfo  {journal} {Nature Communications}\ }\textbf {\bibinfo {volume} {9}},\ \bibinfo {pages} {3975} (\bibinfo {year} {2018})}\BibitemShut {NoStop}%
\bibitem [{\citenamefont {Modic}\ \emph {et~al.}(2021)\citenamefont {Modic}, \citenamefont {McDonald}, \citenamefont {Ruff}, \citenamefont {Bachmann}, \citenamefont {Lai}, \citenamefont {Palmstrom}, \citenamefont {Graf}, \citenamefont {Chan}, \citenamefont {Balakirev}, \citenamefont {Betts}, \citenamefont {Boebinger}, \citenamefont {Schmidt}, \citenamefont {Lawler}, \citenamefont {Sokolov}, \citenamefont {Moll}, \citenamefont {Ramshaw},\ and\ \citenamefont {Shekhter}}]{Modic2021}%
  \BibitemOpen
  \bibfield  {author} {\bibinfo {author} {\bibfnamefont {K.~A.}\ \bibnamefont {Modic}}, \bibinfo {author} {\bibfnamefont {R.~D.}\ \bibnamefont {McDonald}}, \bibinfo {author} {\bibfnamefont {J.~P.}\ \bibnamefont {Ruff}}, \bibinfo {author} {\bibfnamefont {M.~D.}\ \bibnamefont {Bachmann}}, \bibinfo {author} {\bibfnamefont {Y.}~\bibnamefont {Lai}}, \bibinfo {author} {\bibfnamefont {J.~C.}\ \bibnamefont {Palmstrom}}, \bibinfo {author} {\bibfnamefont {D.}~\bibnamefont {Graf}}, \bibinfo {author} {\bibfnamefont {M.~K.}\ \bibnamefont {Chan}}, \bibinfo {author} {\bibfnamefont {F.~F.}\ \bibnamefont {Balakirev}}, \bibinfo {author} {\bibfnamefont {J.~B.}\ \bibnamefont {Betts}}, \bibinfo {author} {\bibfnamefont {G.~S.}\ \bibnamefont {Boebinger}}, \bibinfo {author} {\bibfnamefont {M.}~\bibnamefont {Schmidt}}, \bibinfo {author} {\bibfnamefont {M.~J.}\ \bibnamefont {Lawler}}, \bibinfo {author} {\bibfnamefont {D.~A.}\ \bibnamefont {Sokolov}}, \bibinfo {author} {\bibfnamefont {P.~J.}\ \bibnamefont {Moll}}, \bibinfo {author} {\bibfnamefont {B.~J.}\ \bibnamefont {Ramshaw}}, \ and\ \bibinfo {author} {\bibfnamefont {A.}~\bibnamefont {Shekhter}},\ }\href {\doibase 10.1038/s41567-020-1028-0} {\bibfield  {journal} {\bibinfo  {journal} {Nature Physics}\ }\textbf {\bibinfo {volume} {17}},\ \bibinfo {pages} {240} (\bibinfo {year} {2021})}\BibitemShut {NoStop}%
\bibitem [{\citenamefont {Pocs}\ \emph {et~al.}(2021)\citenamefont {Pocs}, \citenamefont {Siegfried}, \citenamefont {Xing}, \citenamefont {Sefat}, \citenamefont {Hermele}, \citenamefont {Normand},\ and\ \citenamefont {Lee}}]{Pocs2021}%
  \BibitemOpen
  \bibfield  {author} {\bibinfo {author} {\bibfnamefont {C.~A.}\ \bibnamefont {Pocs}}, \bibinfo {author} {\bibfnamefont {P.~E.}\ \bibnamefont {Siegfried}}, \bibinfo {author} {\bibfnamefont {J.}~\bibnamefont {Xing}}, \bibinfo {author} {\bibfnamefont {A.~S.}\ \bibnamefont {Sefat}}, \bibinfo {author} {\bibfnamefont {M.}~\bibnamefont {Hermele}}, \bibinfo {author} {\bibfnamefont {B.}~\bibnamefont {Normand}}, \ and\ \bibinfo {author} {\bibfnamefont {M.}~\bibnamefont {Lee}},\ }\href {\doibase 10.1103/PhysRevResearch.3.043202} {\bibfield  {journal} {\bibinfo  {journal} {Physical Review Research}\ }\textbf {\bibinfo {volume} {3}},\ \bibinfo {pages} {043202} (\bibinfo {year} {2021})}\BibitemShut {NoStop}%
\bibitem [{\citenamefont {Mumford}\ \emph {et~al.}(2021)\citenamefont {Mumford}, \citenamefont {Paul},\ and\ \citenamefont {Kapitulnik}}]{Mumford2021}%
  \BibitemOpen
  \bibfield  {author} {\bibinfo {author} {\bibfnamefont {S.}~\bibnamefont {Mumford}}, \bibinfo {author} {\bibfnamefont {T.}~\bibnamefont {Paul}}, \ and\ \bibinfo {author} {\bibfnamefont {A.}~\bibnamefont {Kapitulnik}},\ }\href@noop {} {\bibfield  {journal} {\bibinfo  {journal} {Phys. Rev. Materials}\ }\textbf {\bibinfo {volume} {5}},\ \bibinfo {pages} {125201} (\bibinfo {year} {2021})}\BibitemShut {NoStop}%
\bibitem [{\citenamefont {Jang}\ \emph {et~al.}(2011)\citenamefont {Jang}, \citenamefont {Ferguson}, \citenamefont {Vakaryuk}, \citenamefont {Budakian}, \citenamefont {Chung}, \citenamefont {Goldbart},\ and\ \citenamefont {Maeno}}]{Maeno2011}%
  \BibitemOpen
  \bibfield  {author} {\bibinfo {author} {\bibfnamefont {J.}~\bibnamefont {Jang}}, \bibinfo {author} {\bibfnamefont {D.~G.}\ \bibnamefont {Ferguson}}, \bibinfo {author} {\bibfnamefont {V.}~\bibnamefont {Vakaryuk}}, \bibinfo {author} {\bibfnamefont {R.}~\bibnamefont {Budakian}}, \bibinfo {author} {\bibfnamefont {S.~B.}\ \bibnamefont {Chung}}, \bibinfo {author} {\bibfnamefont {P.~M.}\ \bibnamefont {Goldbart}}, \ and\ \bibinfo {author} {\bibfnamefont {Y.}~\bibnamefont {Maeno}},\ }\href@noop {} {\bibfield  {journal} {\bibinfo  {journal} {Science}\ }\textbf {\bibinfo {volume} {331}},\ \bibinfo {pages} {186} (\bibinfo {year} {2011})}\BibitemShut {NoStop}%
\bibitem [{\citenamefont {Callen}(1985)}]{Callen1985}%
  \BibitemOpen
  \bibfield  {author} {\bibinfo {author} {\bibfnamefont {H.~B.}\ \bibnamefont {Callen}},\ }\href@noop {} {\emph {\bibinfo {title} {Thermodynamics and an introduction to thermostatistics}}}\ (\bibinfo  {publisher} {Wiley \& Sons},\ \bibinfo {year} {1985})\BibitemShut {NoStop}%
\bibitem [{\citenamefont {Negele}\ and\ \citenamefont {Orland}(1988)}]{Negele1988}%
  \BibitemOpen
  \bibfield  {author} {\bibinfo {author} {\bibfnamefont {J.}~\bibnamefont {Negele}}\ and\ \bibinfo {author} {\bibfnamefont {H.}~\bibnamefont {Orland}},\ }\href {https://books.google.com/books?id=EV8sAAAAYAAJ} {\emph {\bibinfo {title} {Quantum Many-particle Systems}}},\ Advanced Book Classics\ (\bibinfo  {publisher} {Basic Books},\ \bibinfo {year} {1988})\BibitemShut {NoStop}%
\bibitem [{\citenamefont {Onsager}(1931{\natexlab{a}})}]{Onsager1930}%
  \BibitemOpen
  \bibfield  {author} {\bibinfo {author} {\bibfnamefont {L.}~\bibnamefont {Onsager}},\ }\href@noop {} {\bibfield  {journal} {\bibinfo  {journal} {Physical Review}\ }\textbf {\bibinfo {volume} {37}},\ \bibinfo {pages} {405} (\bibinfo {year} {1931}{\natexlab{a}})}\BibitemShut {NoStop}%
\bibitem [{\citenamefont {Onsager}(1931{\natexlab{b}})}]{Onsager1931}%
  \BibitemOpen
  \bibfield  {author} {\bibinfo {author} {\bibfnamefont {L.}~\bibnamefont {Onsager}},\ }\href@noop {} {\bibfield  {journal} {\bibinfo  {journal} {Physical Review}\ }\textbf {\bibinfo {volume} {38}},\ \bibinfo {pages} {2265} (\bibinfo {year} {1931}{\natexlab{b}})}\BibitemShut {NoStop}%
\bibitem [{\citenamefont {Tinkham}(1996)}]{Tinkham1996}%
  \BibitemOpen
  \bibfield  {author} {\bibinfo {author} {\bibfnamefont {M.}~\bibnamefont {Tinkham}},\ }\href {https://books.google.com/books?id=XP\_uAAAAMAAJ} {\emph {\bibinfo {title} {Introduction to Superconductivity}}},\ International series in pure and applied physics\ (\bibinfo  {publisher} {McGraw Hill},\ \bibinfo {year} {1996})\BibitemShut {NoStop}%
\bibitem [{\citenamefont {Zener}(1937{\natexlab{a}})}]{Zener-I-1937}%
  \BibitemOpen
  \bibfield  {author} {\bibinfo {author} {\bibfnamefont {C.}~\bibnamefont {Zener}},\ }\href@noop {} {\bibfield  {journal} {\bibinfo  {journal} {Physical Review}\ }\textbf {\bibinfo {volume} {52}},\ \bibinfo {pages} {230} (\bibinfo {year} {1937}{\natexlab{a}})}\BibitemShut {NoStop}%
\bibitem [{\citenamefont {Zener}(1937{\natexlab{b}})}]{Zener-II-1937}%
  \BibitemOpen
  \bibfield  {author} {\bibinfo {author} {\bibfnamefont {C.}~\bibnamefont {Zener}},\ }\href@noop {} {\bibfield  {journal} {\bibinfo  {journal} {Physical Review}\ }\textbf {\bibinfo {volume} {53}},\ \bibinfo {pages} {90} (\bibinfo {year} {1937}{\natexlab{b}})}\BibitemShut {NoStop}%
\bibitem [{\citenamefont {Chadwick}\ and\ \citenamefont {Liao}(2008)}]{Chadwick2008}%
  \BibitemOpen
  \bibfield  {author} {\bibinfo {author} {\bibfnamefont {R.~S.}\ \bibnamefont {Chadwick}}\ and\ \bibinfo {author} {\bibfnamefont {Z.}~\bibnamefont {Liao}},\ }\href@noop {} {\bibfield  {journal} {\bibinfo  {journal} {SIAM Review}\ }\textbf {\bibinfo {volume} {50}},\ \bibinfo {pages} {313} (\bibinfo {year} {2008})}\BibitemShut {NoStop}%
\bibitem [{\citenamefont {Maali}\ \emph {et~al.}(2005)\citenamefont {Maali}, \citenamefont {Hurth}, \citenamefont {Boisgard}, \citenamefont {Jai}, \citenamefont {Cohen-Bouhacina},\ and\ \citenamefont {Aime}}]{Maali2005}%
  \BibitemOpen
  \bibfield  {author} {\bibinfo {author} {\bibfnamefont {A.}~\bibnamefont {Maali}}, \bibinfo {author} {\bibfnamefont {C.}~\bibnamefont {Hurth}}, \bibinfo {author} {\bibfnamefont {R.}~\bibnamefont {Boisgard}}, \bibinfo {author} {\bibfnamefont {C.}~\bibnamefont {Jai}}, \bibinfo {author} {\bibfnamefont {T.}~\bibnamefont {Cohen-Bouhacina}}, \ and\ \bibinfo {author} {\bibfnamefont {J.-P.}\ \bibnamefont {Aime}},\ }\href {\doibase 10.1063/1.1873060} {\bibfield  {journal} {\bibinfo  {journal} {Journal of Applied Physics}\ }\textbf {\bibinfo {volume} {97}},\ \bibinfo {pages} {074907} (\bibinfo {year} {2005})}\BibitemShut {NoStop}%
\bibitem [{\citenamefont {Keesom}\ and\ \citenamefont {MacWood}(1938)}]{Keesom1938}%
  \BibitemOpen
  \bibfield  {author} {\bibinfo {author} {\bibfnamefont {W.~H.}\ \bibnamefont {Keesom}}\ and\ \bibinfo {author} {\bibfnamefont {G.~E.}\ \bibnamefont {MacWood}},\ }\href {\doibase 10.1016/s0031-8914(38)80195-6} {\bibfield  {journal} {\bibinfo  {journal} {Physica}\ }\textbf {\bibinfo {volume} {5}},\ \bibinfo {pages} {737} (\bibinfo {year} {1938})}\BibitemShut {NoStop}%
\bibitem [{\citenamefont {Landau}\ and\ \citenamefont {Lifshitz}(1976)}]{Landau-Mechanics1976}%
  \BibitemOpen
  \bibfield  {author} {\bibinfo {author} {\bibfnamefont {L.}~\bibnamefont {Landau}}\ and\ \bibinfo {author} {\bibfnamefont {E.}~\bibnamefont {Lifshitz}},\ }\href {https://books.google.com/books?id=e-xASAehg1sC} {\emph {\bibinfo {title} {Mechanics: Volume 1}}},\ Course of theoretical physics\ (\bibinfo  {publisher} {Elsevier Science},\ \bibinfo {year} {1976})\BibitemShut {NoStop}%
\bibitem [{\citenamefont {Landau}\ \emph {et~al.}(1986)\citenamefont {Landau}, \citenamefont {Lifshitz}, \citenamefont {Kosevich},\ and\ \citenamefont {Pitaevskii}}]{Landau-Elasticity1986}%
  \BibitemOpen
  \bibfield  {author} {\bibinfo {author} {\bibfnamefont {L.}~\bibnamefont {Landau}}, \bibinfo {author} {\bibfnamefont {E.}~\bibnamefont {Lifshitz}}, \bibinfo {author} {\bibfnamefont {A.}~\bibnamefont {Kosevich}}, \ and\ \bibinfo {author} {\bibfnamefont {L.}~\bibnamefont {Pitaevskii}},\ }\href {https://books.google.com/books?id=tpY-VkwCkAIC} {\emph {\bibinfo {title} {Theory of Elasticity: Volume 7}}},\ Course of theoretical physics\ (\bibinfo  {publisher} {Elsevier Science},\ \bibinfo {year} {1986})\BibitemShut {NoStop}%
\end{thebibliography}

%

\end{document}